\documentclass{aa1}
\usepackage{graphicx}
\usepackage{color}
\usepackage{natbib}

\def\rsun{R_{\odot}}
\def\msun{M_{\odot}}

\begin{document}
\sloppypar
 
\title{Diagnostics of the Early Explosion Phase of a Classical Nova Using Its X-ray
Emission: A Model for the X-ray Outburst of CI Camelopardalis in 1998}

\offprints{kate@iki.rssi.ru}

\author{E. V. Filippova\inst{1*}, M. G. Revnivtsev\inst{1,2}, and A. A. Lutovinov\inst{1}}

\institute{ 
    Space Research Institute, ul. Profsoyuznaya 84/32, Moscow, 117997 Russia
\and 
Max-Planck Institut f\"ur Astrophysik, Karl Schwarzschild Strasse 1, 86740 Garching-bei-M\"unchen, Germany
          }

        \authorrunning{Filippova et al.}
          \titlerunning{ }
\date{Received: 16 June 2008.}

\abstract{ We have computed a spherically symmetric 
model for the interaction of matter ejected during
the outburst of a classical nova 
with the stellar wind from its optical component.This model is used
to describe the intense X-ray outburst (the peak 3-20 keV 
flux was $\sim$ 2 Crab) of the binary system
CI Camelopardalis in 1998. According to our model, 
the stellar wind from the optical component heated by
a strong shock wave produced when matter 
is ejected from the white dwarf as the result of a thermonuclear
explosion on its surface is the emission 
source in the standard X-ray band. Comparison of the calculated
and observed time dependences of the 
mean radiation temperature and luminosity of the binary system
during its outburst has yielded very 
important characteristics of the explosion.We have been 
able to measure the velocity of the ejected 
matter immediately after the onset of the explosion for the first time:
it follows from our model that the ejected 
matter had a velocity of $\sim$ 2700 km/s even on 0.1-0.5 day after
the outburst onset and it flew with such a velocity 
for the first 1-1.5 day under an external force, possibly,
the radiation pressure from the white dwarf. Subsequently, the matter probably became transparent and
began to decelerate. The time dependence of 
the mean radiation temperature at late expansion phases has
allowed us to estimate the mass of the 
ejected matter, $\sim 10^{-7}-10^{-6} \msun$. The mass loss 
rate in the stellar
wind required to explain the observed peak 
luminosity of the binary system during its outburst has been
estimated to be $\dot{M}\sim (1-2) \times 10^{-6}\msun/yr$.
}

\maketitle

\section{Introduction}
\subsection{X-ray Emission During Classical Nova Outbursts}

Explosive thermonuclear burning of matter on the
surface of a white dwarf is generally believed to be responsible
for the phenomenon of classical novae (see,
e.g., the review by Gallagher and Starrfield (1978)).
During a thermonuclear runaway, a large mass accreted
on the white dwarf surface (up to $\sim10^{-4} \msun$,
Prialnik and Kovetz 2005) is ejected and accelerated
to high velocities, $\sim1000 - 4000$ km/s (see, e.g.,
Prialnik 1986; Prialnik and Kovetz 2005).

At the initial evolutionary phases of a nova outburst,
a high mass loss rate of the white dwarf envelope
causes the effective photosphere of the envelope
to increase in size up to $10^{12}-10^{14}$ cm, depending on
the energy band (see, e.g., Kato and Hachisu 1994).
Subsequently, the photospheric radius decreases over
several months, because the envelope gradually becomes
optically thin, until the entire envelope becomes
completely transparent to emission.

At this evolutionary phase of a classical nova,
the on-going hydrogen burning on the surface of
the white dwarf temporarily turns it to the so-called
"supersoft" X-ray source (see, e.g.,MacDonald et al.
1985; Kahabka and van den Heuvel 1997) - the
entire surface of the white dwarf produces a flux that
is approximately equal to the Eddington one. In an
optically thick regime and at typical white dwarf sizes,
$10^9$ cm, the latter corresponds to a blackbody temperature
of about 20-50 eV.

Classical and recurrent (i.e., erupting several
times, e.g., Nova RS Oph) novae can also be
the sources of standard ($\sim 2-10$ keV) and hard ( $>
20$ keV) X-ray emission. The most popular model for
the generation of such emission is the shock model.

In the external shock model, a dense envelope
that rose from the white dwarf surface as a result of
the explosion and that was accelerated to velocities
$U\sim1000-4000$ km/s produces a strong shock
wave in the stellar wind from the companion star.
The shock-heated wind matter is the source of X-ray
emission up to energies 5-80 keV ($h\nu\sim m_pU^2/2$).
For example, the described mechanism is believed
to be responsible for the generation of hard X-ray
emission from the recurrent Nova RS Oph (see, e.g.,
Bode and Kahn 1985; O'Brien et al.1992; Sokoloski
et al.2006).

Evidently, the external shock model is naturally
linked to the existence of a dense medium near the
binary system; otherwise, the emission measure of the
shock-heated plasma ($EM \sim n^2 dV$ ) would not be
enough to provide the necessary (observed) X-ray
luminosity. However, in most cases, the companions
in classical novae are unevolved main-sequence stars
(Bath and Shaviv 1978) whose weak stellar wind is
unable to produce a dense medium near the binary
system needed for the external shock model.

In the internal shock model, the ejected envelope
has a velocity gradient as a result of which the faster
inner layers catch up with the slower outer ones to
produce a shock wave inside the expanding envelope
(Lloyd et al.1992; O'Brien et al.1994; Mukai and
Ishida 2001).The internal shock model was successfully
applied to such classical novae as Nova Herculis
1991 (Lloyd et al.1992) and Nova Velorum 1999
(Mukai and Ishida 2001).

The internal and external shock models are not
mutually exclusive and can make comparable contributions
to the total X-ray emission from the nova
under certain conditions.The most important factor
here is the acceleration law of the ejected envelope
(here, we discuss only the initial evolutionary
phase of the nova) and the density of the stellar wind
from the companion star. It is possible that not all of
the envelope matter is instantaneously accelerated to
the velocities measured at later envelope expansion
phases from the parameters of optical and infrared
emission lines (see, e.g., Hynes et al. 2002; Das et al.
2006). Therefore, the inner layers of the envelope can
have higher outflow velocities and, hence, an internal
shock that affects significantly the observational
manifestations of the nova in X-rays can be generated
inside the ejected envelope irrespective of the
efficiency of the external shock model.

The envelope acceleration process is important not
only in explaining the X-ray emission from the nova,
but it is also fundamentally important in understanding
the physical processes in the burning envelope
of the white dwarf. Thus, for example, the enrichment
of the envelope with white dwarf matter (mainly
carbon and oxygen) is very important for the formation
of a rapidly expanding envelope in actual nova
explosions. The energy release in the white dwarf
envelope through thermonuclear burning computed
in 1D numerical simulations with convective motions
described by the mixing-length theory allows high
envelope expansion velocities to be obtained (see,
e.g., Starrfield et al.1985; Prialnik 1986; Yaron et al.,
2005). However, more detailed 2D and 3D burning
calculations do not allow such high velocities to be
obtained without the assumptions about greatly enhanced
carbon and oxygen abundances in the envelope
(Kercek et al.1998; Glasner et al. 2005, 2007).
Measuring the envelope expansion law would be the
most important observational test of the white dwarf
envelope burning model.

The expansion law of the envelope ejected from the
white dwarf is exceptionally difficult to test, because
this process takes place within the first hours or even
minutes after the onset of explosive thermonuclear
burning, i.e., long before the maximum optical light
of the nova, i.e., in fact, before its actual detection in
astronomical observations. The observations of hard
X-ray emission from several classical novae a few
days after their maximum light in the optical energy
band are not enough for this.

Only recently has a unique opportunity arisen to
test the nova envelope expansion models using the X-ray
emission from one of the most unusual transients
in the X-ray sky, XTE J0421+560/CI Cam, as an
example.

\subsection{XTE J0421+560/CI Cam}

\begin{figure*}[htb]
\hbox{
\includegraphics[width=\columnwidth]{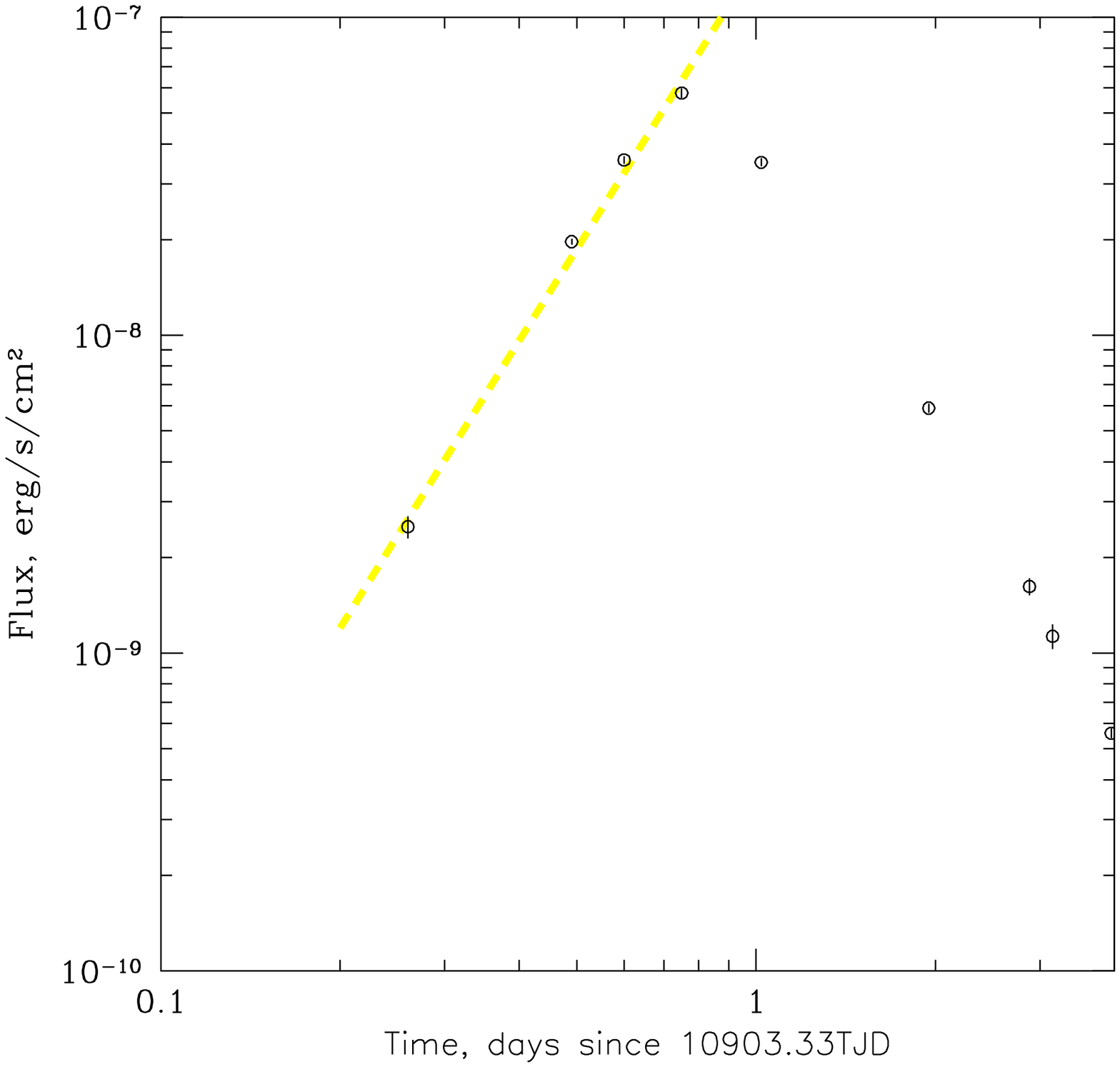}
\includegraphics[width=\columnwidth]{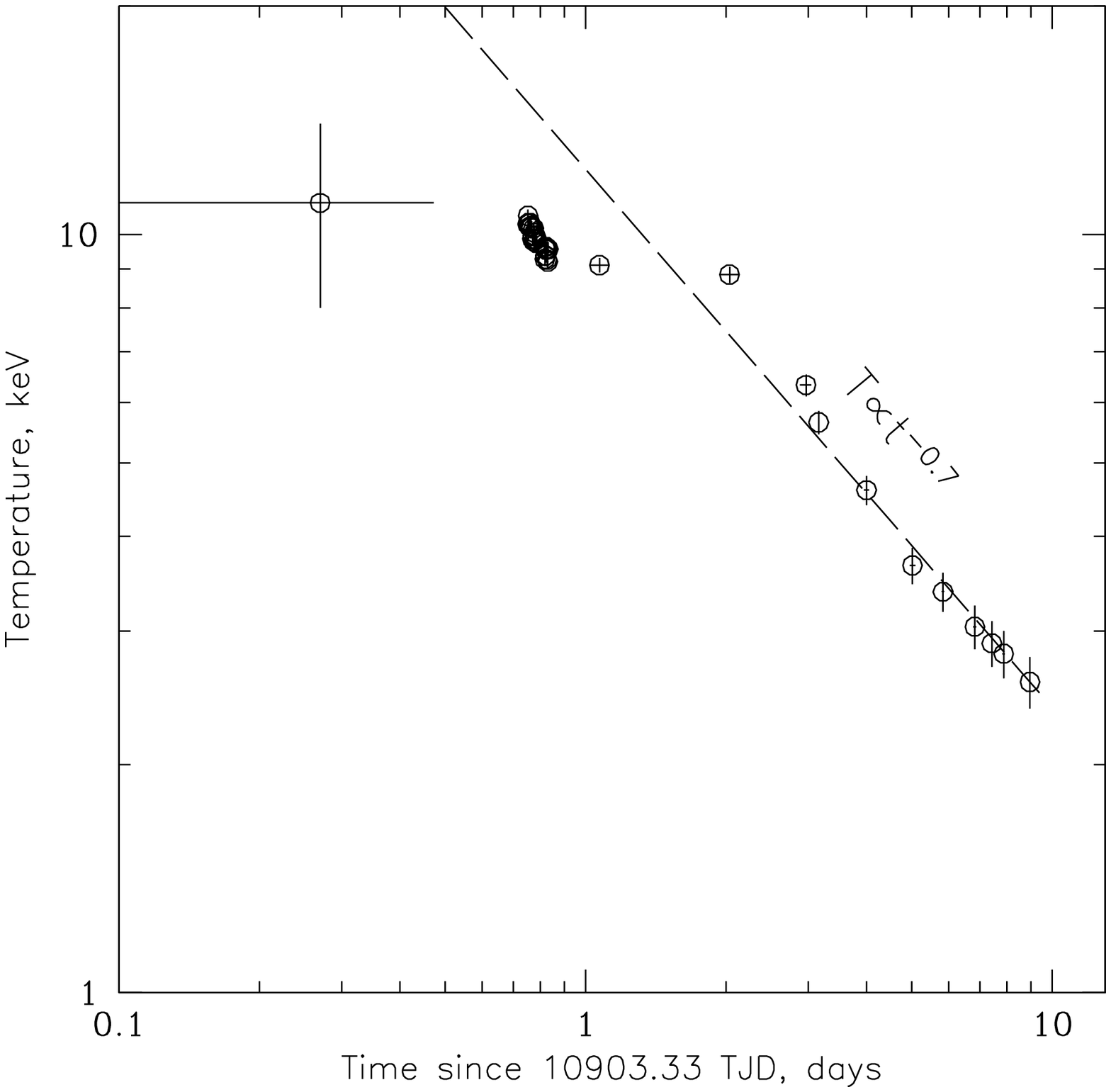}
}\caption{Left: light curve of the system in the 
3-20 keV energy band during the 1998 outburst of CI Cam. 
The dashed straight line
indicates the dependence $L_x \sim t^3$.
Right: time dependence of the mean temperature 
of the emitting matter during the outburst.
The mean temperature was measured from the 
ratio of the 3-5 and 5-20 keV fluxes. 
The first temperature measurement
on 0.1-0.5 day of the outburst 
was made using ASM data (from the ratio 
of the 3-5 and 5-12 keV fluxes); the remaining
measurements were made using PCA/RXTE data. 
The dashed line indicates the dependence $T \sim t^{-0.7}$.}
\end{figure*}

The source XTE J0421+560 (later identified with
the well-known variable star CI Cam; Wagner and
Starrfield 1998) was discovered as an X-ray transient
on March 31, 1998 (Smith et al. 1998). The flux from
the source rose to $\sim$2 Crab in several hours. Subsequent
observations of the system revealed a variable
flux in both optical and radio bands (see, e.g., Clark
et al. 2000).

The properties of XTE J0421+560 differed strikingly
from those of ordinary X-ray transients (see,
e.g., Tanaka and Shibazaki 1996). Various emission
lines typical of the emission from an optically thin
plasma with a temperature of 1-10 keV (Orr et al. 
1998; Ueda et al. 1998; Revnivtsev et al. 1999) but
absolutely atypical of the emission from bright X-ray
transients were detected in its spectrum.

It was also found that, in contrast to ordinary
X-ray transients (accreting neutron stars and black
holes), the flux from XTE J0421+560 was highly stable
at all Fourier frequencies - apart from a general
decline in flux, only an upper limit on the variability of
the source's X-ray flux on time scales of $\sim$20-1000 s
could be obtained (Revnivtsev et al. 1999).

Subsequent studies of the source's behavior at late
evolutionary phases (Ishida et al.2004), in quiescence
(Boirin et al.2002), and investigation of the evolution
of the radio emission from the 1998 outburst
(Mioduszewski and Rupen 2004) suggested that the
intense 1998 outburst was the explosion of a classical
nova in the binary system CI Cam and the compact
object in XTE J0421+560 is most likely an accreting
white dwarf.

For a classical nova, the X-ray emission from
XTE J0421+560 during the 1998 outburst is extraordinarily
intense. For comparison, the peak X-ray flux
from the widely known recurrent Nova RS Ophiuchi
(Sokoloski et al.2006; Bode et al. 2006) was a factor
of 10-30 lower than that from CI Cam during its
1998 outburst. A strong stellar wind from the optical
component in the binary system, a B[e] giant star
(Hynes et al.2002; Barsukova et al.2006; Robinson
et al.2002), is responsible for such a high luminosity
of the CI Cam/XTE J0421+560 outburst (we
will show this below). It is the presence of such a
dense medium around the expanding envelope of the
nova that produced the necessary emission measure
($EM\sim n^2 dV$ ) of the shock-heated plasma.

The assumption that the explosion of a classical
nova is responsible for the giant X-ray outburst of
XTE J0421+560/CI Cam in 1998, along with the
high X-ray intensity of the transient (the peak flux
from the source reached $\sim$2 Crab), which allowed a
large set of high-quality observational data covering
almost the entire evolution of the transient's X-ray
outburst to be obtained, provides an excellent opportunity
to test the most important ingredients of the
external shock model for classical nova explosions.A
unique feature of the set of XTE J0421+560/CI Cam
observations from the standpoint of studying the evolution
of the classical nova envelope expansion is
the transient's light curve obtained by the All-Sky
Monitor (ASM) of the RXTE observatory half a day
before the X-ray flux peak. It allows the motion of the
expanding envelope to be ``seen'' starting from several
hundred white dwarf radii ($\sim10^{11-12}$ cm). The evolution
of the X-ray outburst of XTE J0421+560, which,
on the whole, lasted about 10 days, was well covered
by RXTE observations (Revnivtsev et al. 1999). This
also makes it possible to compare the predictions of
the shock models during the decline in X-ray flux.

This paper is the first in a series of papers that
investigate the behavior of the X-ray emission from
CI Cam using hydrodynamic simulations of the expansion
of the white dwarf envelope and its interaction
with the stellar wind from the companion star.

The goal of this paper is to demonstrate that the
main features of the X-ray outburst of CI Cam can
be explained in terms of the model of a shock wave
passing through the stellar wind during a classical
nova explosion.We investigated in detail the following
important points of the explosion:

\begin{itemize}

\item the envelope motion at the initial expansion
stages after the explosion;
\item the deceleration of the envelope as it interacts
with the stellar wind;
\item the contribution from radiative cooling to the
change in the temperature of the stellar wind from the
companion passed through the shock wave.
\end{itemize}

\subsection{Basic Characteristics of the Emission from CI Cam
During the 1998 Outburst}

During the giant outburst of CI Cam in 1998, the
X-ray emission from this binary system had several
characteristic features that, as we will show below,
can be explained in terms of the model of emission
generation as the shock wave produced by the expanding
envelope of the white dwarf passes through
the stellar wind from the optical companion.

\begin{itemize}

\item At the rise phase of the light curve, the time
dependence of the source's 3-20 keV luminosity was
$L_x \sim t^{\sim 3}$ (Fig.1, left).
\item The peak luminosity of the source was 
$L_{peak}(3-20 keV)\sim 3\times 10^{37}(d/2 kpc)^2$ erg/s, 
where d is the distance to the source.
\item The mean (effective) radiation temperature of
the source was constant and equal to kT $\sim$ 10 keV
for 2 days from the outburst onset (Fig.1, right).
\item After the second day from the outburst onset,
the radiation temperature decreased approximately as
$T \sim t^{-0.7-0.6}$.
\end{itemize}

\section{The Method Of Numerical
Calculations}
\subsection{The Computational Scheme}
In this paper, we used numerical hydrodynamic
calculations to derive the time dependences of the
effective temperature and X-ray luminosity of a classical
nova.

In our calculations, we used a 1D spherically
symmetric code in Lagrangian coordinates with a
staggered mesh (the cell radius, velocity, and mass
are determined at the cell boundaries, while the
density, pressure, and internal energy are determined
at the cell mass centers). A more detailed description of
this computational scheme can be found in Janka
et al. (1993).

We disregarded the gravitational attraction from
the white dwarf, since the shock wave in our calculations
is generated at a distance, $5\times10^{11}$ cm,
at which the escape velocity from the white dwarf,
$\sim$230 km/s, is much lower than the envelope velocity,
$\sim$3000 km/s. We also disregarded the influence
of the optical star on the shock propagation.

\subsection{Radiative Cooling and Thermal Instability}

\begin{figure}[]
\includegraphics[width=0.9\columnwidth,angle=0.,bb=20 150 570 500]{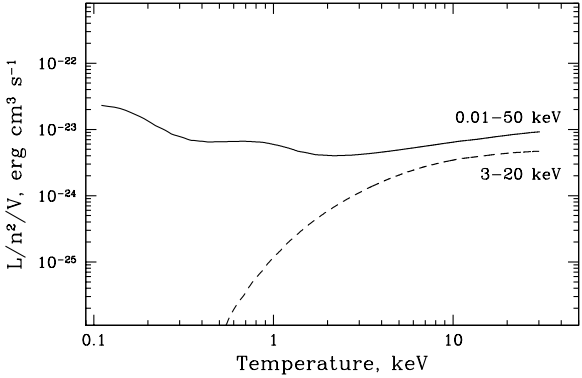}
\caption{Plasma cooling functions in two energy bands 
obtained from the APEC model. The functions were reduced to the total
particle number density.}
\label{cool_fun}
\end{figure}

In our calculations, we took into account the
radiative cooling of the stellar wind matter from the
optical companion passed through the shock wave.
The bremsstrahlung of an optically thin plasma was
assumed to be the main mechanism of the radiative
losses.We calculated the rate of plasma energy
losses through radiation using the APEC model
(http://hea-www.harvard.edu/APEC/REF) by assuming
that ionization equilibrium was established
instantly in the heated matter. The derived cooling
functions are shown in Fig.2. For convenience, these
functions were reduced to the total particle number
density. The cosmic abundances were taken from
Grevesse and Sauval (1998), the mean molecular
weight of the plasma particles at such abundances
is $\mu = 0.61$, and $n/n_e = 1.93$, where $n$ is the total
number of particles. To take into account the radiative
energy losses, we used the 0.01-50 keV energy
band in our calculations, which is sufficient for the
temperatures expected in our model. The 3-20 keV
energy band was used to construct the model light
curves.

A simple estimate of the stellar wind density at
which radiative cooling begins to affect significantly
the thermal balance of the post-shock matter can be
obtained by comparing the radiative cooling time $\tau_{rad}$
through bremsstrahlung alone with the characteristic
times in our problem, several days. For our estimate,
we will take 1 day. The gas in our calculations was
assumed to be ideal with the adiabatic index $\gamma = 5/3$:
$$
\tau_{rad} \sim {3/2 n_1kT \over{\Lambda n_1^2 }}={3/2 kT \over{\Lambda n_1}} s,
$$
where $T$ is the temperature of the emitting plasma, $n_1$
is the total density of the stellar wind matter passed
through the shock wave, and $\Lambda$ is the plasma emissivity
(cooling function). For the observed temperature,
kT $\sim$ 10 keV, the plasma emissivity is 
$\Lambda = 6.45 \times 10^{-24} erg\, cm^3 s^{-1}$ 
and a radiative cooling time of
1 day corresponds to an unperturbed stellar wind
density $n_0 = n_1/4 \sim 10^{10} cm^{-3}$.

In the case where the radiative cooling time of the
shock-heated matter is shorter than the characteristic
time of its expansion (several days), an isobaric
thermal instability develops (see, e.g., Field 1965). In
1D calculations, this leads to a number of nonphysical
effects. However, since the matter cooled below a
temperature of $\sim$1 keV ceases to radiate in the energy
band of interest to us (3-20 keV), its specific temperature
does not affect the results of our calculations
as long as the thermal instability develops at constant
pressure. Therefore, when the cell cooling time was
less than 0.1 day (which corresponds to temperatures
$<$ 0.9 keV at typical densities in our problem), we
switched off the cooling in this cell and, at the same
time, excluded its emission from our calculations of
the observed X-ray luminosity and the mean temperature
of the X-ray emission.

Let us now show that our approximation of
the instant establishment of ionization equilibrium
downstream of the shock is valid. Before estimating
the time it takes for ionization equilibrium to be
established, we must calculate the time it takes for a
Maxwellian ion velocity distribution to be established
and the equipartition time of the post-shock ion and
electron temperatures $\tau_{ie}$, since the shock wave heats
only the ions. Spitzer et al.(1965) showed that
$$
\tau_{ii}={{11.4 A_i^{1/2} T^{1/2}_i}\over{n_i Z_i^4 ln\Lambda}}s
$$

$$
\tau_{ie}=5.87 {{(T_e A_i+T_i A_e)^{3/2}}\over{n_i Z_i^2 ln\Lambda \sqrt{A_e A_i}}}s
$$

where $T_i$ and $T_e$ are the ion and electron temperatures
in kelvins, $n_i$ is the ion density, $A_e = 1/1836$ is the
atomic weight of the electron, $A_i$ is the atomic weight
of the ion, and $ln\Lambda$ is the Coulomb logarithm.

For our estimates, we will take $A_i = 1, ln\Lambda = 15,
T_i = 10 keV, T_e = 1eV, Z_i = 1,$ and $n_i = n_e$. Then,

$$
\tau_{ii}\sim 95 n_{i, 10}^{-1} s
$$

$$
\tau_{ie}\sim 0.04 n_{i, 10}^{-1}s
$$

where $n_{i,10} = n_i/10^{10} cm^{-3}$. 
The time it takes for
the electron temperature to be established 
is $\tau_{ee}=\sqrt{A_e}\tau_{ii}=2.2n_{e, 10}^{-1}$ s.

To estimate the time it takes for ionization equilibrium
to be established downstream of the shock, we
will use a formula from Masai (1994):
$$
\tau_{\rm eq}\sim 10^{12}/n_e \sim 100n_{e, 10}^{-1}s,
$$

where $n_e$ is the electron density.
It follows from our estimates that the longest time
it takes for the temperature to be established 
($\tau_{ii}\sim 50 s$) and the time it takes for ionization equilibrium to
be established ($\tau_{eq} \sim 50 s$) downstream of the shock
are much shorter than the radiative cooling time of
the matter ($\tau_{rad} = 1$ day) at the same density (here,
$n_i \sim n_e \sim n_1/2 = 2 \times 10^{10} cm^{-3}$). 
Hence, our approximation
of the instant establishment of ionization
equilibrium downstream of the shock is valid. We
also see from our estimates that the shock at matter
parameters typical of our problem is not radiative
(since $\tau_{ii} << \tau_{rad}$), i.e., the density jump at the shock is
defined by the formula $n_1={{\gamma+1}\over{\gamma-1}}n_0$ and is $n_1/n_0 =
4$ for our value of $\gamma = 5/3$.

\subsection{Calculating the Mean Temperature of the X-ray
Emission}
\begin{figure}[]
\includegraphics[width=\columnwidth]{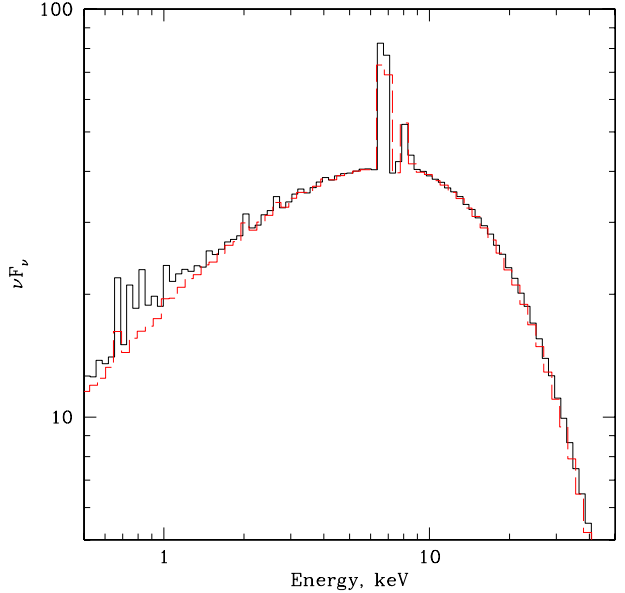}
\caption{Spectrum of the matter passed through the shock
wave composed of the spectra of the individual cells
within which the matter has the same temperature (solid
line). The dashed line indicates the spectrum of a single temperature
plasma with the temperature determined
from the flux ratio (see the text).}\label{composition} 
\end{figure}

Below, we will show that, in general, the post-shock
stellar wind matter is a multi-temperature
plasma. Consequently, the radiation temperature that
we measure based on X-ray observations is an average
and it may not be equal to the temperature at the
shock front. Ttherefore, to obtain the calculated mean
temperature, we used the same averaging procedure
as that for the observations. We calculated the ratio
of the 3-5 and 5-20 keV fluxes, which, in turn,
corresponds to a certain temperature in the radiation
model of a single-temperature optically thin plasma.
The flux ratio in the model of a single-temperature
plasma was calculated using the APEC model. The
efficiency of the method is shown in Fig. 3, where
the solid line indicates the combined spectrum of
the post-shock matter at some instant of time and
the dashed line indicates the spectrum of a single temperature
plasma with the temperature determined
by the method described above. We see from Fig. 3
that the method is efficient in our case.

\subsection{Initial Conditions}

The stellar wind density, temperature, and velocity
profiles were specified in our calculations as the initial
conditions. The first inner cell of the computational
grid with a finite mass was specified as the piston-
envelope. This allowed us to subsequently consider
the influence of the piston envelope mass on the shock
dynamics.

In our calculations with the piston-envelope, the
initial cell size was $\Delta r = 5\times 10^{11}$ cm. The inner
boundary of the computational grid was placed at
$10^{10}$ cm. We performed calculations with various initial
grid cell sizes to ascertain the optimal computational
time/accuracy ratio.An increase in the resolution
by a factor of 10 caused a change in the mean
temperature of the emitting matter by a few percent
while increasing significantly the computational time.

In our case, the stellar wind temperature does not
affect the shock generation; therefore, it was taken to
be approximately equal to the effective temperature of
the optical star, $\sim10000-20000 K: T_0=1$ eV.
 The temperature of the piston-envelope was taken to be
equal to that of the stellar wind.

Since the stellar wind velocity (35 km/s; Robinson
et al.2002) is low compared to the envelope
velocity ($\sim$1000-4000 km/s) and does not affect the
shock generation either, it was set equal to zero in our
calculations for simplicity.

For the models with a constant envelope expansion
velocity, constant velocities were maintained at
both boundaries of the envelope cell.

When the shock wave was modeled in the Sedov
regime (a strong point explosion in a spherically
symmetric medium), the internal energy of the first
four cells increased to the necessary explosion energy,
whereupon the system relaxed. For our calculations
in this regime, we set the initial cell size equal to
$\Delta r = 2\times 10^{11} $cm.

\section{Positions Of The Components
In CI Cam At The Time Of Its Outburst}

\begin{figure}[]
\centerline{
\includegraphics[height=\columnwidth,bb=330 280 865 1090,clip]{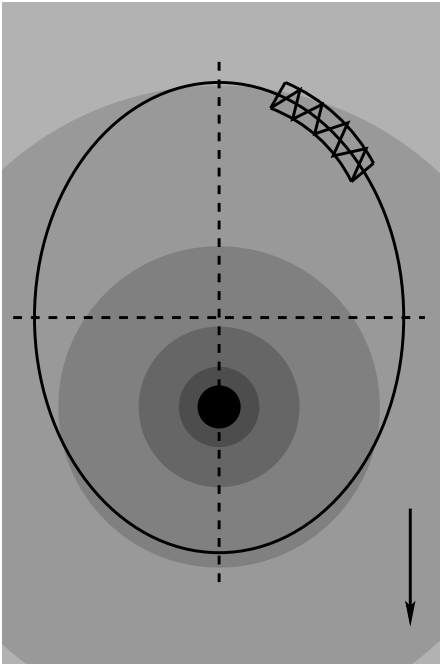}
}
\caption{Relative positions of the stars during the outburst. 
The hatched region corresponds to the possible positions of the
white dwarf during the outburst. 
The black circle marks the optical companion. 
The gray rings indicate the regions with a
dense stellar wind; the wind density at the outer 
boundary of each ring is half that at the inner boundary. Arrow points 
the direction to the observer.}\label{system}
\end{figure}

The optical component in the binary system
CI Cam is an early B[e] star (Hynes et al.2002;
Barsukova et al.2006), presumably a B4 III-V giant
(Barsukova et al. 2006). The radius of the optical
component is 
$R_\ast=3.6-5.9 R_\odot = 2.5 \times 10^{11} - 
4.1 \times 10^{11}$cm. 

If the star is a supergiant (see, e.g., Hynes
et al. 2002; Robinson et al. 2002), then its radius is
considerably larger and can reach $3\times10^{12}$ cm.
The orbital parameters of the binary were determined
by Barsukova et al. (2006): its orbital period
is $P_{orb} = 19.41\pm0.02$ days, the semi-major axis of
its orbit is $a \sin i \sim 70\rsun = 4.8\times10^{12}$ cm 
(here, $i$ is the inclination of the binary), 
and its eccentricity is
$e = 0.62\pm0.07$. The distance to the binary system is
not yet known accurately; in different papers, it varies
between 1 and 10 kpc (Hynes et al.2002; Robinson
et al. 2002; Ishida et al.2004). Barsukova et al. (2006)
found the distance to the binary to be within the range
1.1-1.9 kpc. In this paper, we took 2 kpc.

Using the time of periastron passage 
$T_0 =52198.5$ MJD (Barsukova et al. 2006) and the outburst
onset time 50903.33 MJD, we calculated the
orbital phase of the binary, $\phi_{orb} \sim 0.273$, 
and the separation between its 
components, $r \sin i \sim 6.6 \times 10^{12}$ cm, 
at the outburst time. The measurement
errors of the orbital period and eccentricity lead to
inaccuracies in determining the orbital phase and,
accordingly, the separation between the components.
In Fig.4, the region of possible positions of the white
dwarf at the outburst time is hatched and corresponds
to the following parameters: 
$r \sin i \sim (5.4-7.5) \times 10^{12}$ cm 
and $\phi_{orb} = 0.2-0.34$. The position of the
binary relative to the line of sight was found by
Barsukova et al. (2006) based on the shape of the
radial velocity curve.

\section{Stellar Wind From The Optical
Component}

\subsection{Estimating the Photoabsorption in the Stellar Wind
from the Optical Component}

\begin{figure}[]
\centerline{
\includegraphics[width=0.6\columnwidth,angle=-90,bb=76 32 579 720,clip]{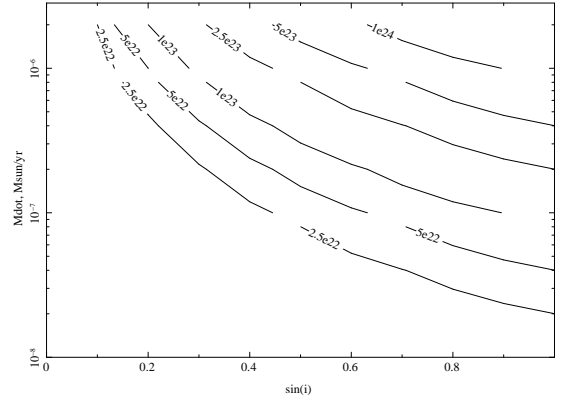}
}
\caption{Equivalent column for photoabsorption as a function of 
binary inclination and mass loss rate by the optical 
companion.}\label{nh_contours}
\end{figure}

Robinson et al.(2002) estimated the mass loss
rate in the stellar wind from the optical component in
CI Cam to be $\dot{M} >10^{-6} \msun/yr$. 
Using the stellar
mass loss rate in the wind, we can estimate the stellar
wind density near the white dwarf and compare it
with the photoabsorption in the X-ray spectrum of
XTE J0421+560.

Significant photoabsorption was detected in the
X-ray emission from CI Cam/XTE J0421+560 in
various observing periods (Revnivtsev et al. 1999;
Boirin et al. 2002). The emission from XTE J0421+560
in quiescence (Boirin et al. 2002) most likely results
from mass accretion onto the white dwarf.
The equivalent column for photoabsorption in this
spectrum may be the sum of the photoabsorptions
in the stellar wind within the binary system and
in the accretion column in the immediate vicinity
of the white dwarf. In this case, the large column
density, $N_HL \sim 5\times 10^{23} cm^{-2}$, 
obtained by Boirin et al. (2002) does not allow us to unequivocally
attribute this to the contribution from the absorption
precisely in the stellar wind.

To estimate the stellar wind density, we can use the
column for photoabsorption at the beginning or at the
peak of the light curve for the 1998 outburst 
($N_HL \sim 5\times10^{22} cm^{-2}$). 
In this case, the X-ray emission
emerges far from the white dwarf surface and, hence,
it is in any case free from this additional photoabsorption.

To calculate the dependence of the column density
on the mass loss rate by the optical component, we
assumed that the stellar wind from the optical star
was spherically symmetric about the star, had a velocity
$v_{wind} = 35$ km/s, and its density $n$ depended
on the distance to the stellar center $R$ as
$$
n = {\dot{M}\over{4 \pi r^2 v_{\rm wind}\mu m_p}}=1.4\times10^{11} r_{12}^{-2}v_{35}^{-1}\dot{M}_{-7},
$$
where $\mu$ is the mean molecular weight of a single
particle, $m_p$ is the proton mass, $r_{12} = r/10^{12}$ cm,
$\dot{M}_{-7}=\dot{M}/10^{-7} \msun/yr$, 
and $v_{35} = v/35$ km/s. 
The relative positions of the star and the line of sight with
respect to the binary system were determined above.

In Fig. 5, the contours indicate several photoabsorption
columns for various orbital inclinations and
mass loss rates by the optical component. We see
from the figure that the outflow rate of the stellar wind
from the optical component at the observed 
$N_HL =5\times10^{22} cm^{-2}$ can lie in the range 
$4\times10^{-8} < \dot{M} < 2\times10^{-6} \msun/yr$, 
while $N_HL = 5\times10^{23} cm^{-2}$ gives
a constraint $\dot{M} > \times10^{-7} \msun/yr$. 
These estimates
are consistent with the values from Robinson
et al. (2002).

\subsection{Estimating the Density Distribution of the Stellar
Wind Relative to the White Dwarf}

\begin{figure}[]
\includegraphics[width=0.9\columnwidth,angle=0,bb=20 145 570 500]{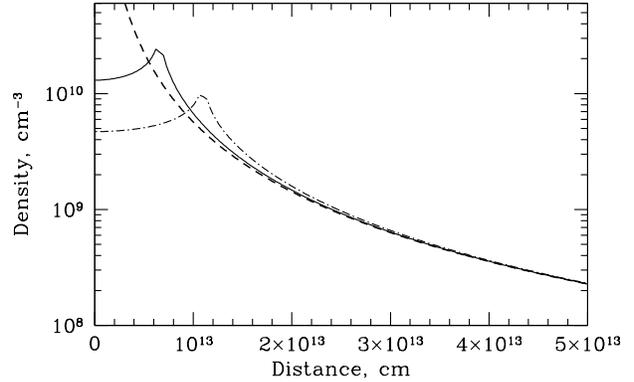}
\caption{Averaged stellar wind density profiles: 
for $\sin i = 1$ (solid line) 
and $\sin i = 0.6$ (dash-dotted line). The dashed
curve indicates the law of decrease $n \propto r^{-2}$. 
In this calculation, the mass loss rate 
by the optical star in the wind is
$\dot{M}=4\times10^{-7}\msun/yr$.}\label{dens_prof}
\end{figure}

The stellar wind density at large distances $r$ from
the star generally decreases as $n\propto r^{-2}$. 
However, since at the outburst onset the envelope that rose
from the white dwarf generates a shock wave around
the latter even inside the binary system, we cannot
assume the density profile  $n\propto r^{-2}$  (where $r$ is the
distance from the white dwarf) to be correct from the
very onset of the outburst.

At small distances from the white dwarf, $r < r_c$,
the stellar wind density will not decreases as $r^{-2}$
(where $r$ is the distance from the white dwarf) but
will be more uniform. At larger distances, $r > r_c$, the
transition to a decreasing density $n\propto r^{-2}$ will occur.

To make the first simplest estimates of the density
distribution of the interstellar medium as a function of
the distance from the white dwarf, we used assumptions
about the stellar wind motion similar to those
made in the previous section and calculated the mean
stellar wind density in the volumes between concentric
spheres centered at the location of the white
dwarf during the outburst. The radius of the optical
component in our calculations was taken to be $5.9\rsun$.
The derived average stellar wind density profile as a
function of the distance from the white dwarf is shown
in Fig. 6. We see that the density is almost constant
near the white dwarf and behaves as a typical stellar
wind density profile with  $n\propto r^{-2}$ starting from some
distance. We will use this approximation of the density
profile in our calculations. It also follows from the
derived profile that $r_c \sin i \sim 6.6\times10^{12}$ cm. 
Below, $r_c$
denotes the distance at which a constant stellar wind
density profile transforms into a decreasing one.

In Fig. 7, the contours show how the stellar wind
density near the white dwarf (at $r < r_c$) depends on
the orbital inclination and the mass loss rate by the
optical component.
\begin{figure}[]
\centerline{
\includegraphics[width=0.6\columnwidth,angle=-90,bb=76 32 579 720,clip]{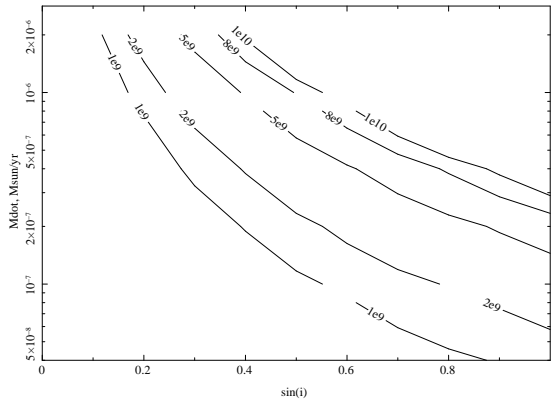}
}
\caption{Stellar wind density near the white dwarf for
various mass loss rates by the optical companion and
various orbital inclinations.}\label{dens_const_cont}
\end{figure}

\section{The Scheme Of Shock Waves
During  The Outburst}

\begin{figure}[]
\centerline{
\includegraphics[width=0.8\columnwidth,bb=0 0 140 160,clip]{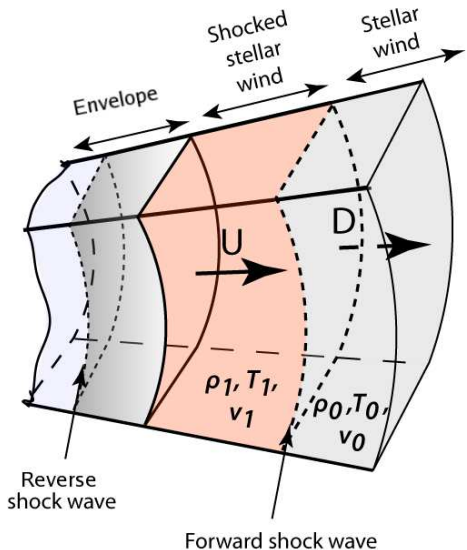}
}
\caption{Scheme of shock waves generated by the expanding envelope: 
$U$ is the velocity of the contact discontinuity or the
envelope and $D$ is the velocity of the forward shock. 
The subscripts ``0'' and ``1'' denote the quantities in the unperturbed stellar
wind and downstream of the forward shock, respectively.}\label{scheme}
\end{figure}

The external shock model that we consider here
suggests that the white dwarf envelope with a mass
$M_{ej}$ moves under the action of an external force (possibly,
the radiation pressure from the white dwarf on
which thermonuclear burning of the remaining accreted
matter continues) for some time $\Delta t$ after the
onset of its expansion; subsequently, it flies ballistically,
i.e., the law of motion of the envelope is determined
only by its interaction with the surrounding
matter and is not the result of its acceleration by the
pressure from the burning part of the envelope.

As a result of the envelope expansion, an external
shock travels through the stellar wind, while either
a shock travels through the envelope inward or a
rarefaction wave is initially generated which subsequently
transforms into a shock as the envelope expands
and as the pressure in it decreases, depending
on the envelope velocity for a given ratio of the initial
pressures.

A schematic view of the system of shocks is shown
in Fig. 8. In principle, the X-ray emission can come
not only from the forward shock but also from the
reverse one. However, since the reverse shock should
be weaker than the forward one and since the radiative
cooling of the matter downstream of the reverse
shock is strong due to its high density, the expected
temperature of the matter downstream of the reverse
shock will not be high enough for it to radiate in the
X-ray band under consideration (3-20 keV, corresponding
to temperatures $T \sim 10^7-10^8 K$). Therefore,
we assume here that the contribution from the
envelope to the observed X-ray emission is negligible
and disregard it. In our calculations, the envelope
was considered as a piston of finite mass producing
a shock in the stellar wind and the letter $U$ denotes
the velocity of the piston-envelope.

\section{The Effect Of Nonuniformities
In The Stellar Wind Density Profile
On The Mean Temperature
And Luminosity Of The Emitting
Matter}

\begin{figure}[]
\centerline{
\includegraphics[width=0.9\columnwidth,angle=0,bb=20 145 570 500 ]{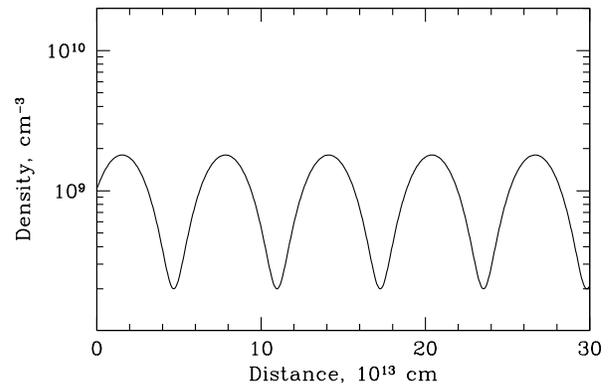}
}
\caption{Density profile of the medium at the initial time:
$n_0 = 10^9 + 8\times10^8 \sin r_{13}$, where $r_{13}$ is the radius in
$10^{13}$ cm. }\label{dens_sinr}
\end{figure}

The properties of the X-ray emission from CI Cam
during its outburst strongly suggest that it is formed
by the emission of an ordinary hot optically thin
plasma (Ueda et al.1998; Orr et al.1998; Revnivtsev
et al.1999). The luminosity of a unit volume of
optically thin hot plasma is proportional to the square
of its density and, hence, is very sensitive to density
variations around the white dwarf. In the binary
system CI Cam, which consists of a young star with
a strong stellar wind and an accreting white dwarf,
there must exist various deviations from a spherically
symmetric distribution of stellar wind matter around
the white dwarf, such as the density jumps in the wake
produced when the white dwarf moves through the
stellar wind (Dumm et al.2000), the nonspherically
symmetric stellar wind from the optical star, etc. This
was clearly demonstrated in a recent paper devoted
to numerical simulations of the matter distribution in
the recurrent Nova RS Oph, which also contains an
optical star with a strong stellar wind (Walder et al.
2008).

To estimate the effect of stellar wind density
nonuniformities on the mean radiation temperature
and luminosity, we calculated the envelope expansion
with a constant velocity in a medium with
a variable density that depended on the radius as
$n_0 = 10^9 + 8\times10^8 \sin(r_{13})$, where $r_{13}$ is the radius
in $10^{13}$ cm (see Fig.9 ). For simplicity, the radiative
losses of the post-shock matter were disregarded in
the calculations.

The envelope motion with a constant velocity
through a medium with a constant density produces a
shock wave with a constant temperature at the front;
consequently, the mean temperature of the postshock
matter also remains constant (if the radiative
losses are small). The luminosity of the shock-heated
matter depends on the time as $L\propto t^3$ (i.e., it is
proportional to the volume of the emitting matter; a
more detailed derivation of the formula will be given
in the Section ``Rise Phase of the Light Curve'').

For a variable (changing by an order of magnitude)
pre-shock matter density, the mean post-shock
temperature and the luminosity of the heated matter
will behave in a different way (Figs.10 (top left) and 10 (top right)).
Figures 10 (bottom left) and 10 (bottom right) show deviations of the mean
temperature from the expected constant value and
deviations of the luminosity from the expected law
$L \propto t^3$, respectively. 
We clearly see from Fig. 10 that
the mean radiation temperature is much less affected
by density jumps: for deviations of the luminosity
from the law $L \propto t^3$ by a factor of 12, the radiation
temperature changes only by a factor of 1.5. Therefore,
to qualitatively describe the evolution of the X-ray
emission from the 1998 outburst of CI Cam, we
will primarily rely on the observed behavior of the
emitting-plasma temperature.

\begin{figure*}[]
\centerline{
\hbox{
\includegraphics[width=0.9\columnwidth,angle=0.,bb=20 145 570 500]{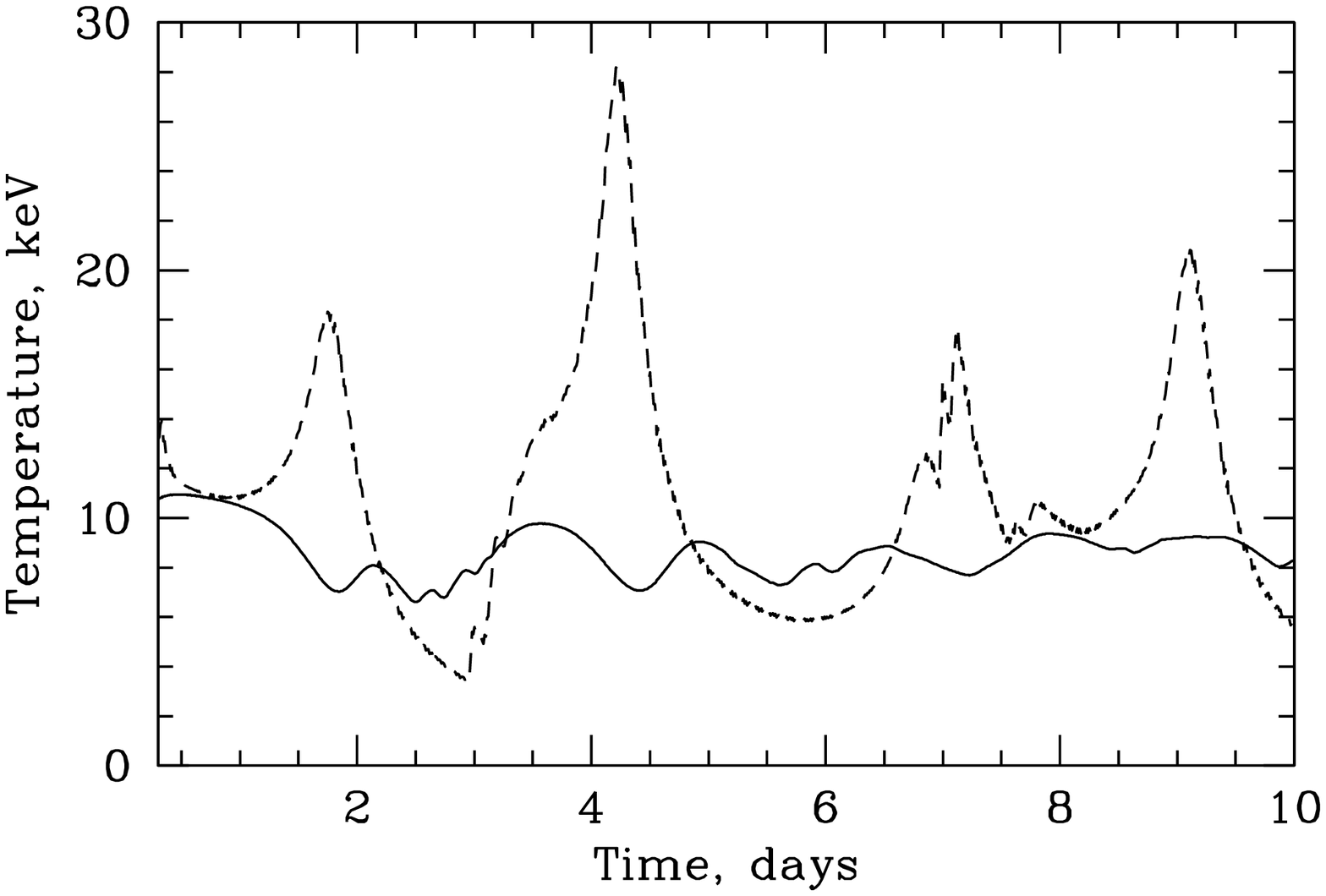}
\includegraphics[width=0.9\columnwidth,angle=0.,bb=20 145 570 500]{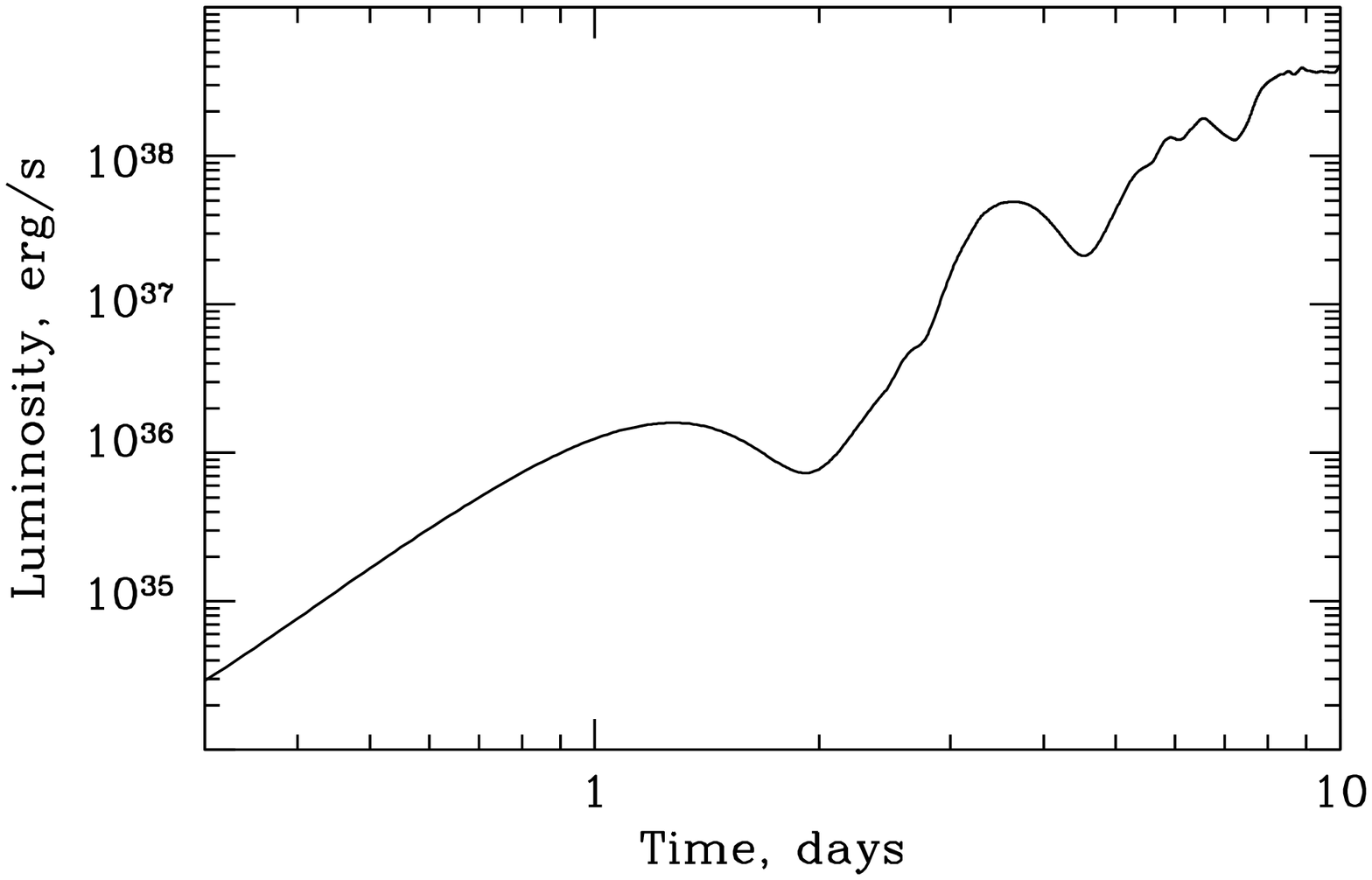}
}
}
\centerline{
\hbox{
\includegraphics[width=0.9\columnwidth,angle=0.,bb=20 145 570 500]{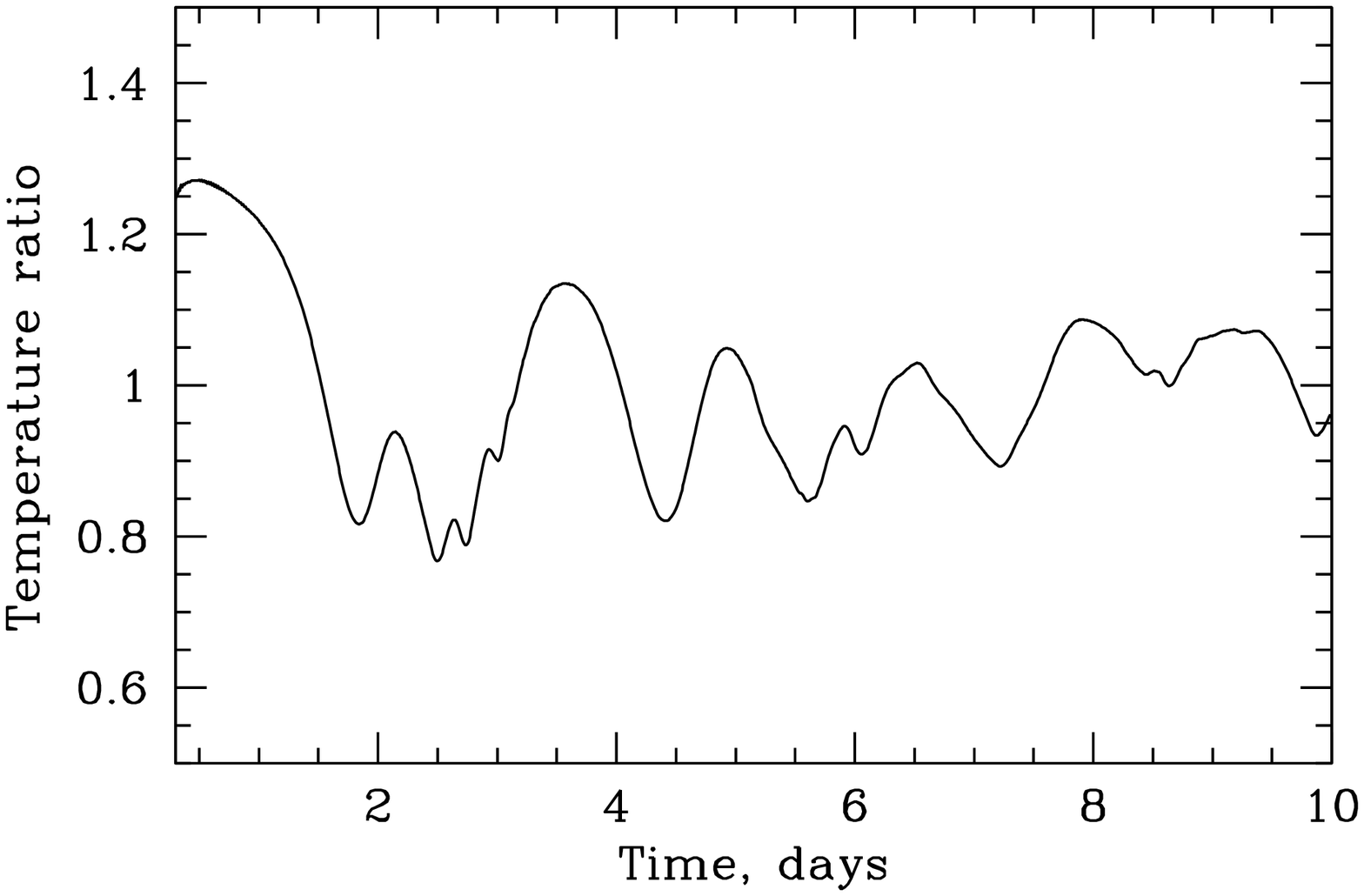}
\includegraphics[width=0.9\columnwidth,angle=0.,bb=20 145 570 500]{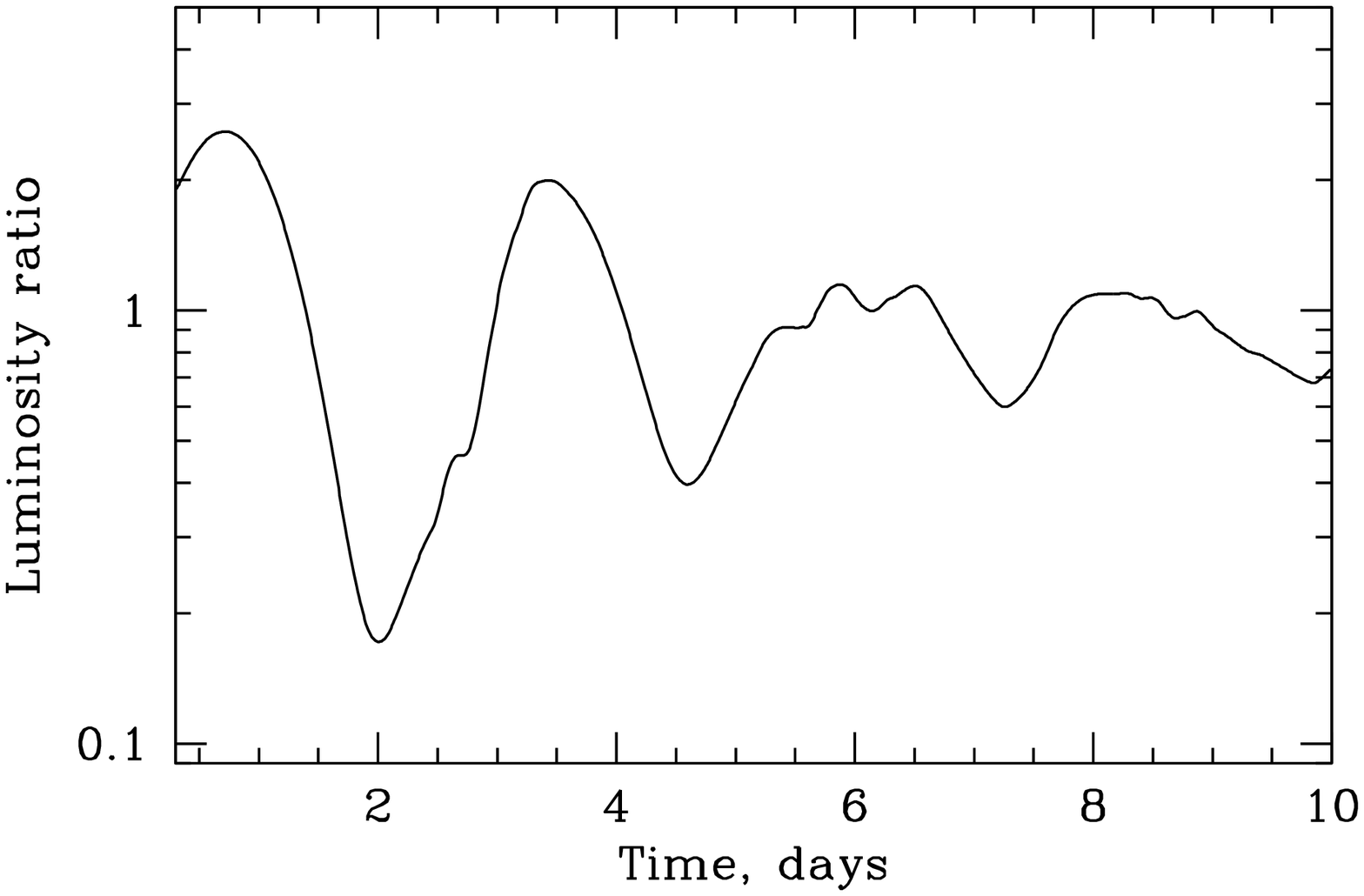}
}
}
\caption{Left top: time dependence of the mean matter 
temperature (solid line) and the post-shock temperature (dashed line) when
the piston moves with a constant velocity 
in a medium with a stellar wind density 
profile $n\sim \sin r$; right top: time dependence of the
luminosity; left bottom: deviation of the mean 
matter temperature from the mean value; 
right bottom: deviation of the plasma luminosity in this
calculation from the law $L = At^3$ 
expected if the pre-shock matter were distributed uniformly.}\label{if_ro_sin}
\end{figure*}

\section{Mean Radiation Temperatures
And Shock Velocity In The Initial
Expansion Period}

\begin{figure*}[]
\centerline{
\hbox{
\includegraphics[width=0.9\columnwidth,angle=0.,bb=20 145 570 500]{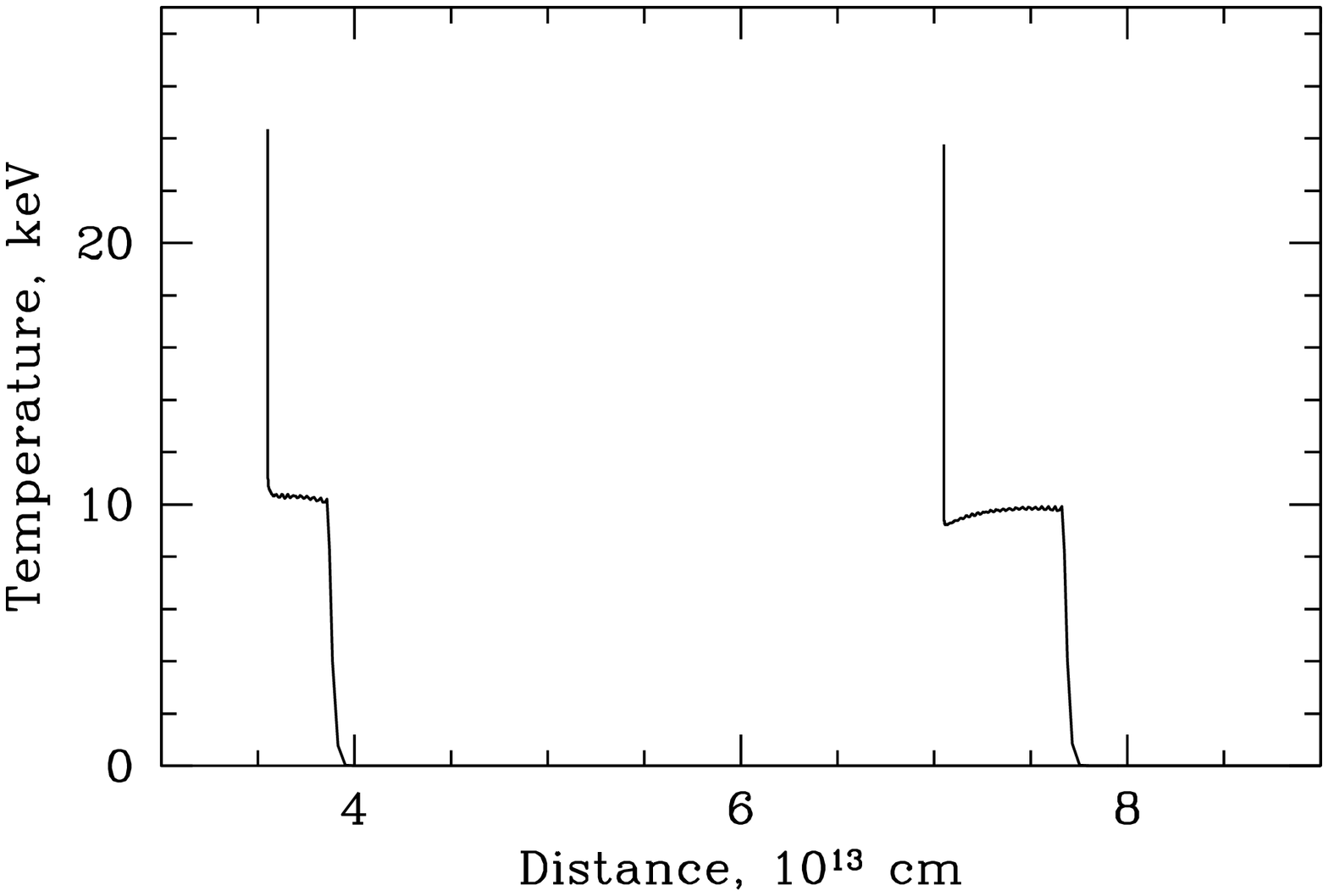}
\includegraphics[width=0.9\columnwidth,angle=0.,bb=20 145 570 500]{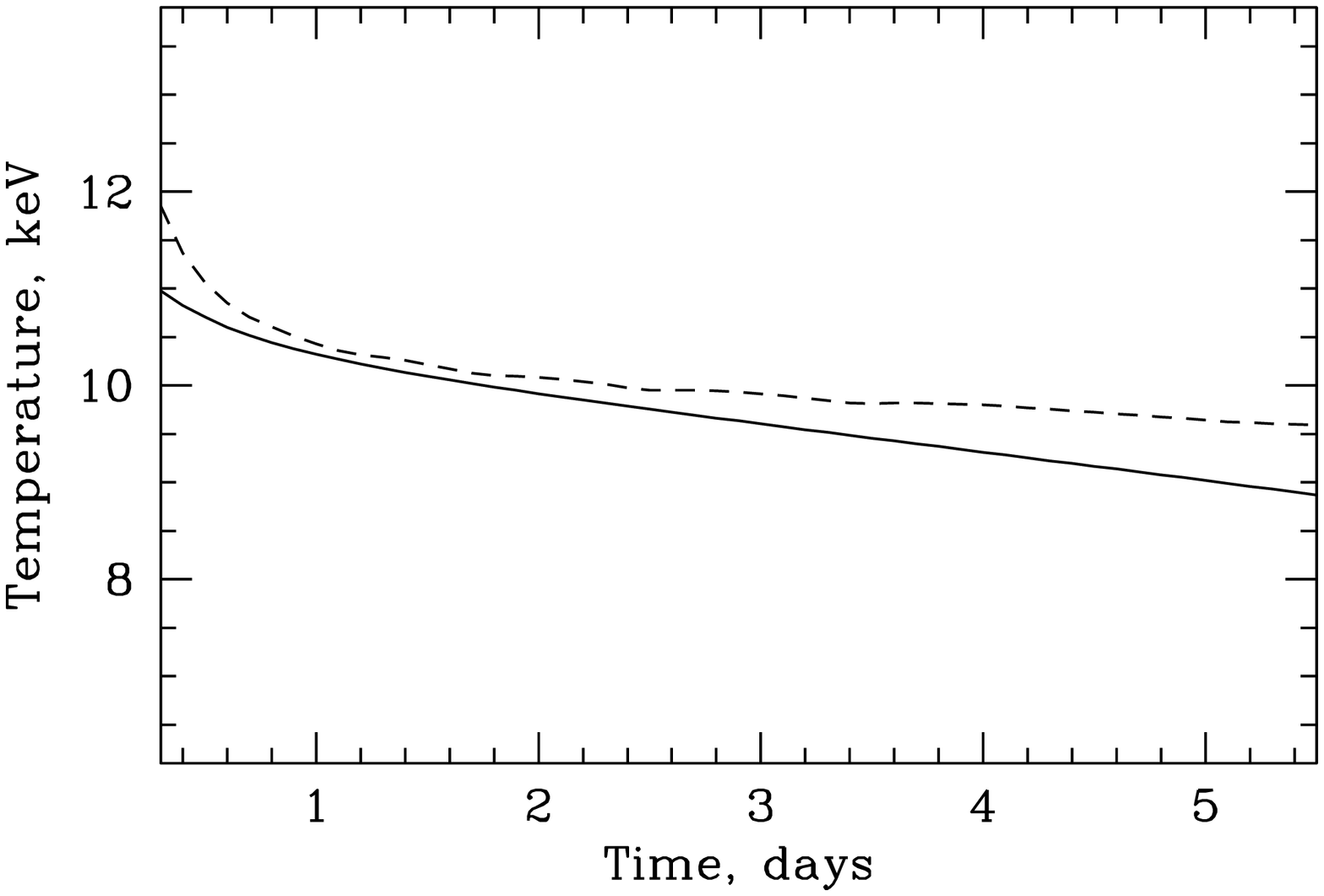}
}
}
\caption{Left: post-shock temperature profiles when the envelope 
moves with a constant velocity through a medium with a
constant density $n = 10^9 cm^{-3}$ 1.5 and 3 days after the onset 
of envelope expansion; right: time dependences of the mean
temperature (solid line) and the temperature at the shock 
(dashed line).}\label{temp_roconst_uconst}
\end{figure*}

Starting from the first hours of outburst development,
the ASM/RXTE flux measurements for
CI Cam in several channels allow the temperature
of the emitting plasma to be estimated. We used the
ASM/RXTE hard X-ray measurements of CI Cam
in the 3-5 and 5-12 keV energy channels, where the
effect of photoabsorption clearly seen in the source's
spectrum on 0.7 day (Revnivtsev et al. 1999) is
weaker. Using the known hardness (flux ratio) of
the Crab Nebula in the ASM 3-5 and 5-12 keV
energy channels, we estimated what hardness of
CI Cam corresponded to the radiation temperature
of an optically thin plasma (the APEC model) within
the first 0.5 day after the outburst onset. When
modeling the radiation of an optically thin plasma,
we also assumed the presence of photoabsorption
in the source's spectrum with a column 
$N_HL =5\times 10^{22} cm^{-2}$. 
Since PCA/RXTE observations are
available after 0.7 day from the outburst onset, we
used the temperature measured precisely by PCA in
this outburst period.

We clearly see from the ASM and PCA data that
the radiation temperature during the first two days remained
approximately constant and equal to $\sim 10$ keV
(Fig.1).

In the case of a strong shock wave, the matter
temperature at the front is essentially determined only
by its velocity:
$$
kT_1 = {2\mu m_p D^2 } {\gamma -1 \over{(\gamma+1)^2}},
$$

where $\gamma$ is the adiabatic index of the interstellar gas
and $D$ is the shock velocity. Consequently, the shock
velocity can be estimated from the measurements of
the radiation temperature at the shock front:

$$
D=(\gamma+1)\sqrt{{kT_1}\over{2(\gamma-1)\mu m_p}}=2894 (T_{1,10keV})^{1/2}km/s,
$$

where $T_{1,10keV}=T_{1,keV}/10$. 

However, in our X-ray observations, we measure
not the plasma temperature at the shock front but the
flux-averaged temperature of the entire post-shock
matter. The relationship between the mean temperature
and the temperature at the shock front depends
on the distribution of the pre-shock stellar wind density
and the radiative cooling of the post-shock heated
matter. Let us consider how these factors affect the
relationship between the temperatures.

\subsection{Shock Motion Through a Region with a Uniform
Stellar Wind Density}

When the piston moves with a constant velocity
through a stellar wind with a constant density,
the post-shock temperature is constant and depends
weakly on time if the radiative cooling of the matter
is negligible. Figure 11(left) shows the post-shock temperature
profiles 1.5 and 3 days after the explosion\footnote{The 
temperature rise near the piston envelope is a numerical
effect (see, e.g., Noh 1978); this is the so-called ``entropy
wake'' (Samarsky and Popov 1992).}
for an initial density of the medium $n = 10^9 cm^{-3}$.
The radiative cooling of the matter at such  density is
negligible, but a reduction in the temperature near the
inner boundary due to radiative losses is already noticeable
on day 3. The time dependences of the mean
post-shock temperature (solid line) and the temperature
at the shock (dashed line) in this calculation are
shown in Fig. 11(right). The theoretical temperature at the
shock is 10.5 keV. We see from Fig. 11 that the mean
temperature within the first several days does not differ
greatly from the temperature at the shock and, to a
first approximation, the observed temperature can be
set equal to the temperature at the shock to estimate
the shock velocity.

For the observed temperature $kT \sim 10$ keV in the
case of negligible radiative losses, we obtain the velocity
of the forward shock in the initial expansion
period (until $\sim$2 days from the explosion onset), 
$D \sim 2900$ km/s.

In turn, the velocity of the strong shock ahead of
a spherical piston in a homogeneous medium is determined
by the piston velocity as $D \sim 1.1U$ (Sedov
1945). Consequently, the envelope velocity in our case
is $U \sim 2630$ km/s.

\subsection{Shock Motion Through a Region with a Stellar
Wind Density Profile $\sim r^{-2}$}

\begin{figure*}[]
\centerline{
\hbox{
\includegraphics[width=0.9\columnwidth,angle=0.,bb=20 145 570 500]{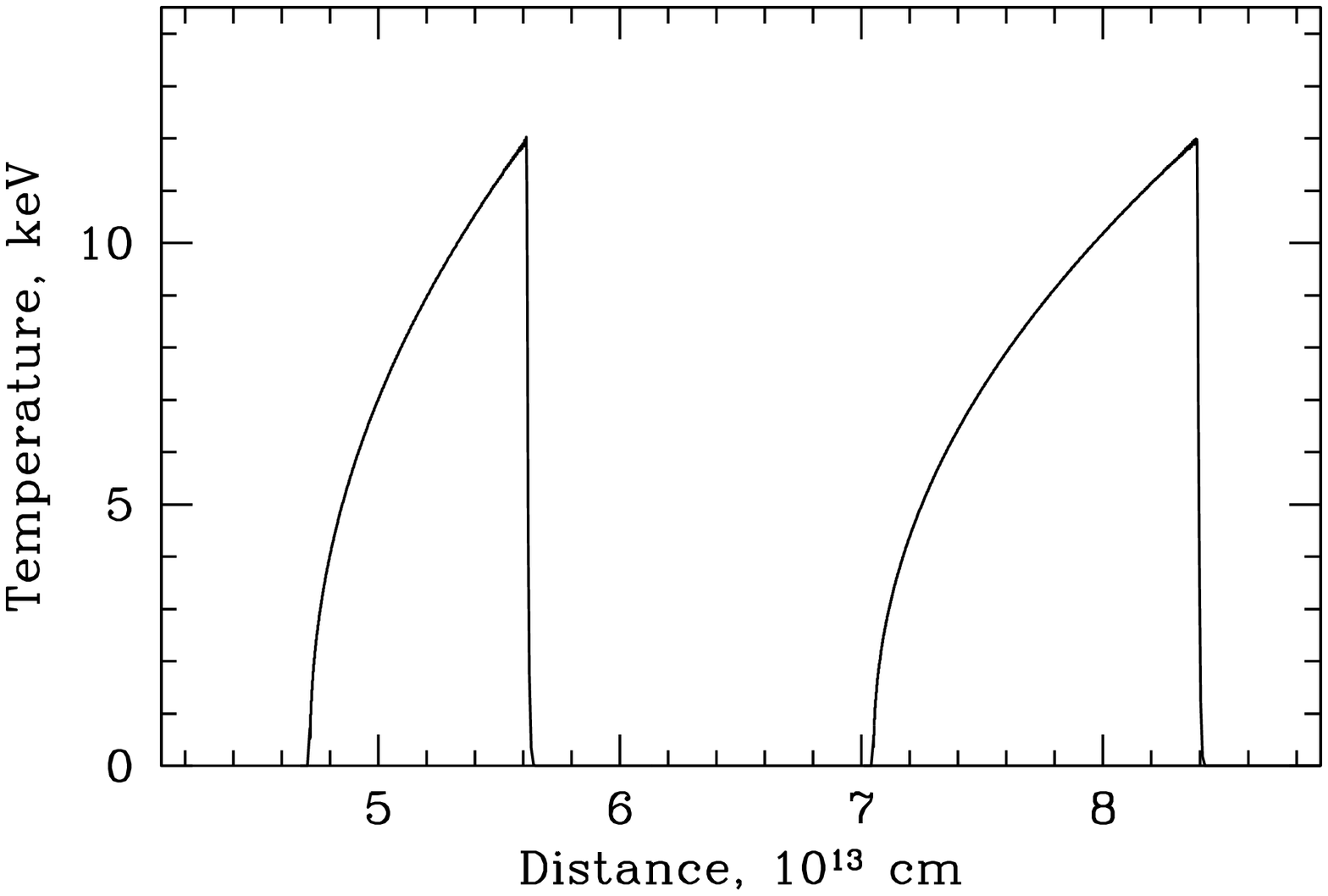}
\includegraphics[width=0.9\columnwidth,angle=0.,bb=20 145 570 500]{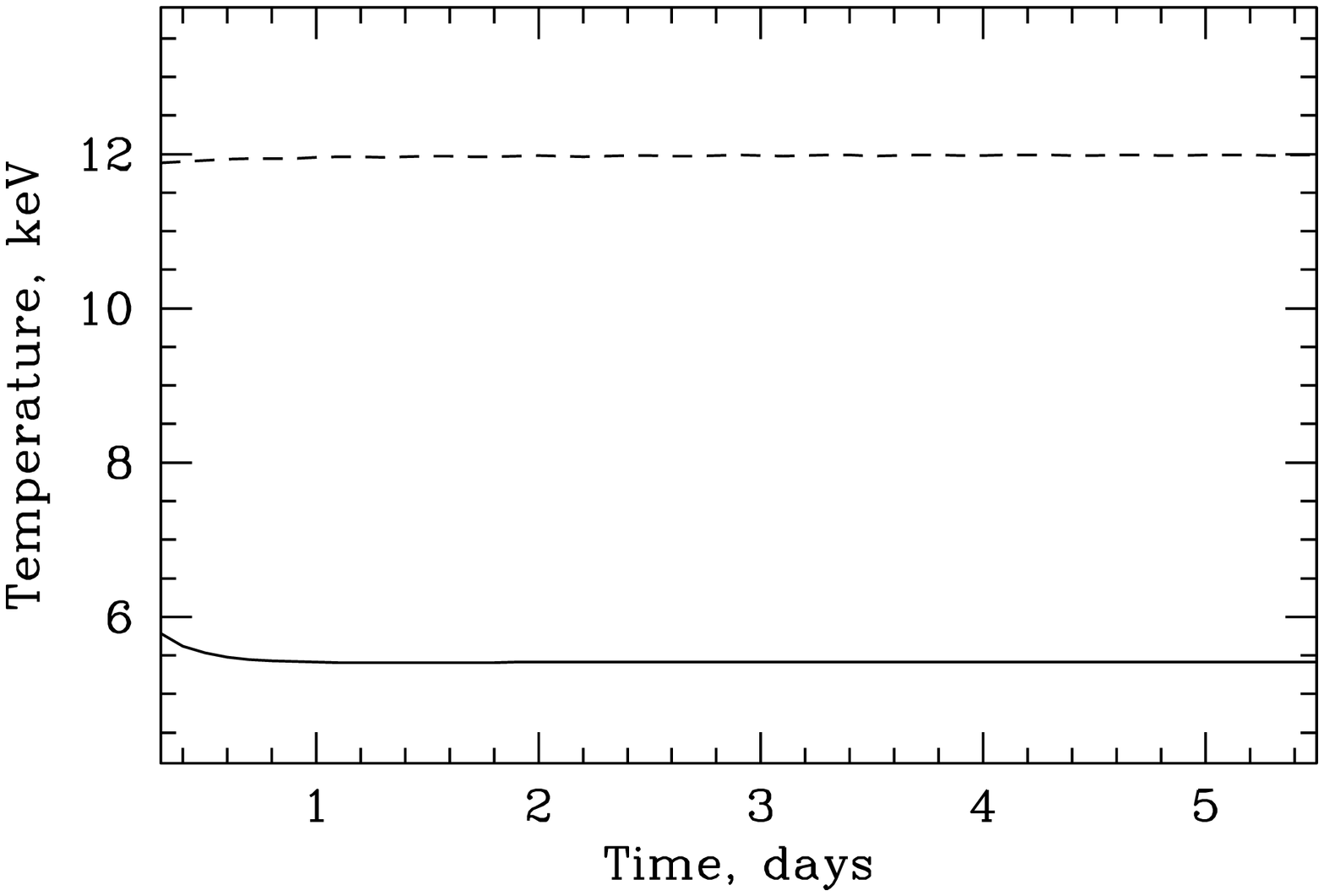}
}
}
\caption{Left: post-shock temperature profiles when 
the envelope moves with a constant velocity through a 
medium with
a density $n \sim r^{-2}$ 1.5 and 3 days after the explosion; 
right: time dependences of the mean temperature (solid line) and the
temperature at the shock (dashed line). }\label{temp_ror2_uconst}
\end{figure*}

When the shock moves with a constant velocity
through a medium with a stellar wind density profile
($n_0 = 5\times10^{10}(r/5.1\times10^{11})^{-2} cm^{-3}$), 
the postshock
temperature profiles 1.5 and 3 days after the
onset of envelope expansion have the form shown in
Fig.12(left). The time dependences of the mean temperature
and the temperature at the shock are shown in
Fig.12(right).

Although the mean temperature in this case remains
constant and we could determine the velocity
of the piston envelope using the observed mean radiation
temperature, other observed quantities (e.g.,
the absolute luminosity of the heated matter and its
time dependence) in this model disagree strongly with
observations.The light curve for this stellar wind
density profile is shown in Fig. 13; we see that the luminosity
begins to decline almost immediately, while
it follows from observations that it rises within the first
$\sim$0.75 day.

It follows from this calculation that the stellar wind
density profile near the white dwarf should differ from
the law $n \sim r^{-2}$.
\begin{figure}[]
\centerline{
\includegraphics[width=0.9\columnwidth,angle=0,bb=20 145 570 500]{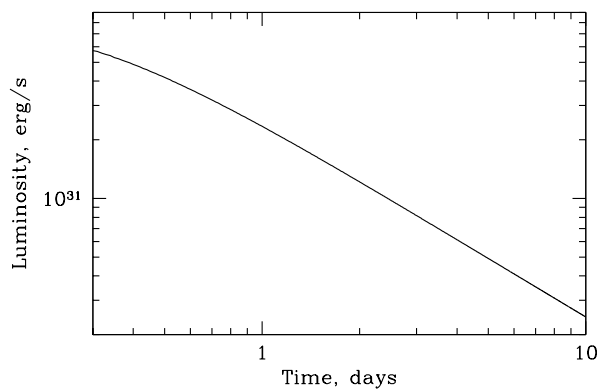}
}
\caption{Time dependence of the luminosity when the shock moves 
with a constant velocity through matter with a density
profile $n\sim r^{-2}$.}\label{lum_ror2_uconst}
\end{figure}

\subsection{The Effect of Radiative Cooling on the Relationship
Between the Mean Temperature and the Temperature at the Shock}

\begin{figure*}[]
\centerline{
\hbox{
\includegraphics[width=0.9\columnwidth,angle=0.,bb=20 145 570 500]{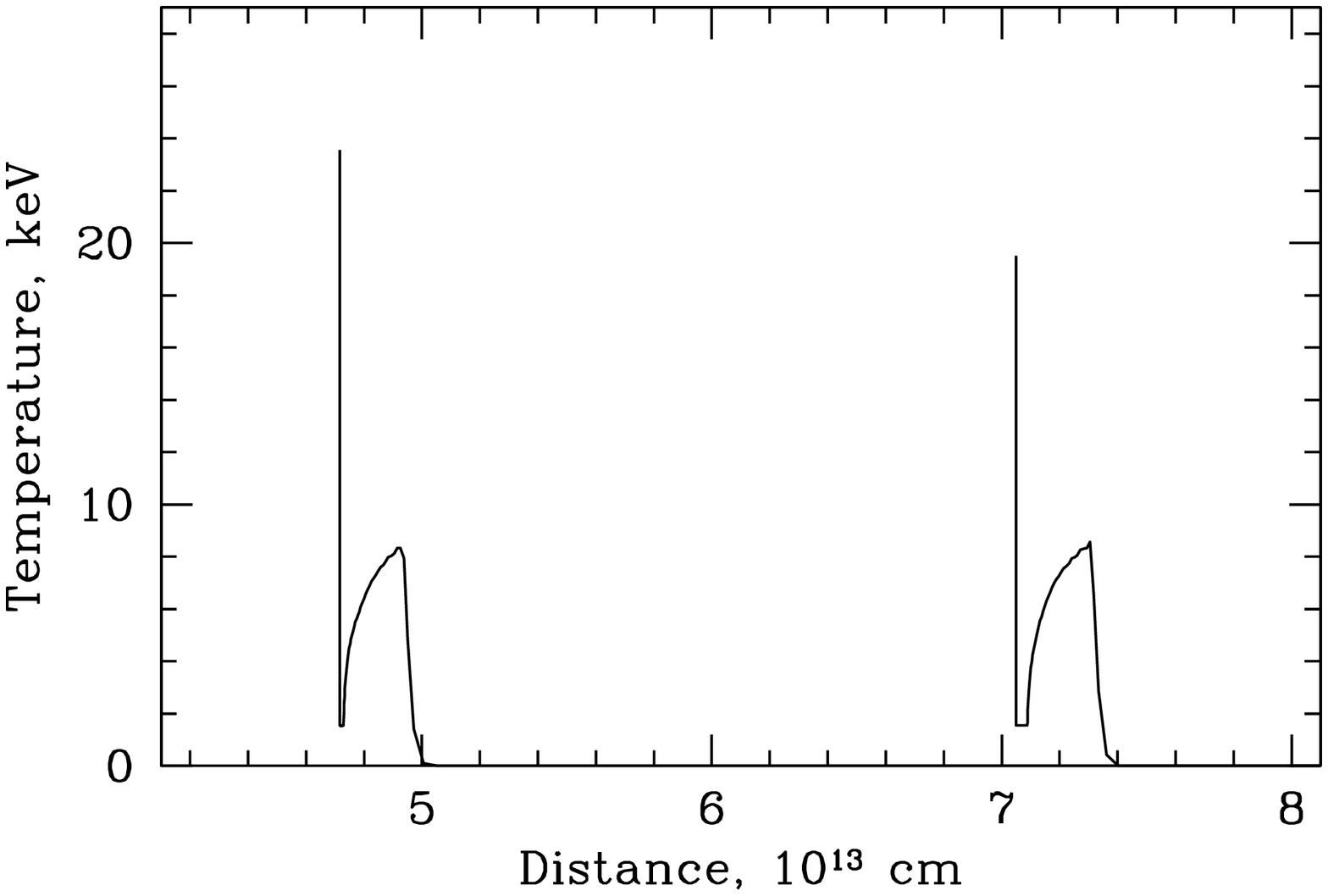}
\includegraphics[width=0.9\columnwidth,angle=0.,bb=20 145 570 500]{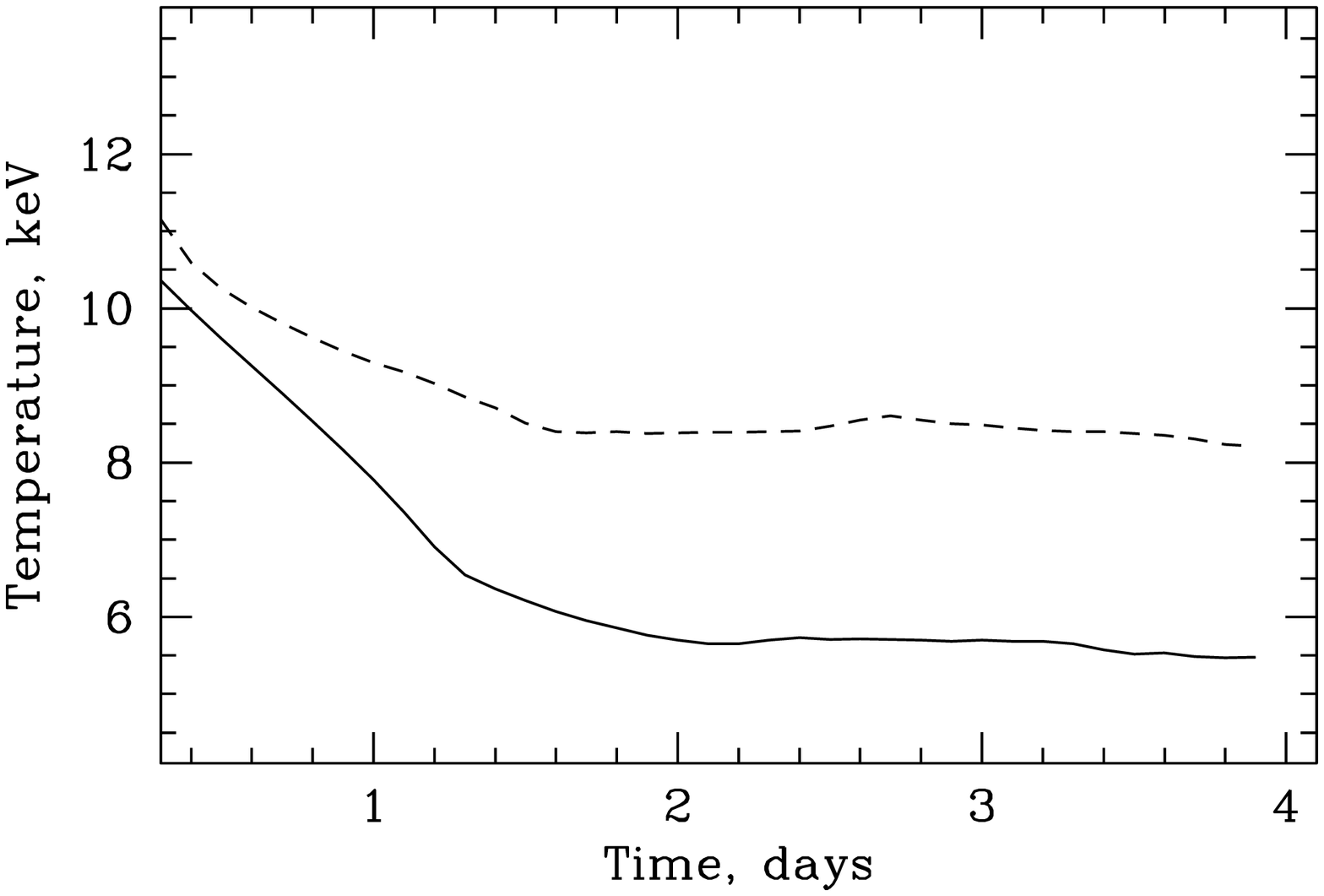}
}
}
\caption{Left: post-shock density profiles when the 
envelope moves with a constant velocity through a medium with a constant
density $n = 10^{10} cm^{-3}$ 2 and 3 days after 
the onset of envelope expansion. A shelf formed by the 
cells with switched-off
radiative cooling is seen in the temperature profile 
on the third day. Right: time dependences of the mean temperature (solid line)
and the temperature at the shock (dashed line).}\label{temp_roconst_uconst_wr}
\end{figure*}

It follows from Fig. 7 that the stellar wind density
near the white dwarf can be $n_0 \sim 10^{10} cm^{-3}$. 
At such densities, radiative cooling plays a significant role: it
reduces the shock velocity and leads to the growth of
thermal instability in the matter passed through the
shock.

When the envelope moves with a constant velocity
through a medium with a constant density 
$n_0 = 10^{10} cm^{-3}$, the post-shock temperature profiles two
and three days after the onset of envelope expansion
have the form shown in Fig. 14(left). A region with
switched-off radiative cooling in its cells (a shelf) is
seen in the temperature profile on the third day near
the inner boundary. Figure 14(right) shows the time dependences
of the post-shock temperature (dashed line)
and the mean temperature (solid line).It follows from
this figure that in the case of intense radiative cooling
of the post-shock matter, its mean temperature
decreases greatly even at a constant piston velocity.
This behavior of the mean temperature is in conflict
with the observed one, from which it follows that
the temperature of the emitting matter was almost
constant for the first two days (see Fig. 1).

Consequently, we can constrain the stellar wind
density near the white dwarf, $n_0 < 10^{10} cm^{-3}$, 
by assuming that the envelope is not accelerated. 
Interestingly,
the derived constraint on the stellar wind
density near the white dwarf agrees well with the
density estimated from the source's luminosity (see
below).
\begin{figure}
\includegraphics[width=0.9\columnwidth,angle=0.,bb=20 145 570 500]{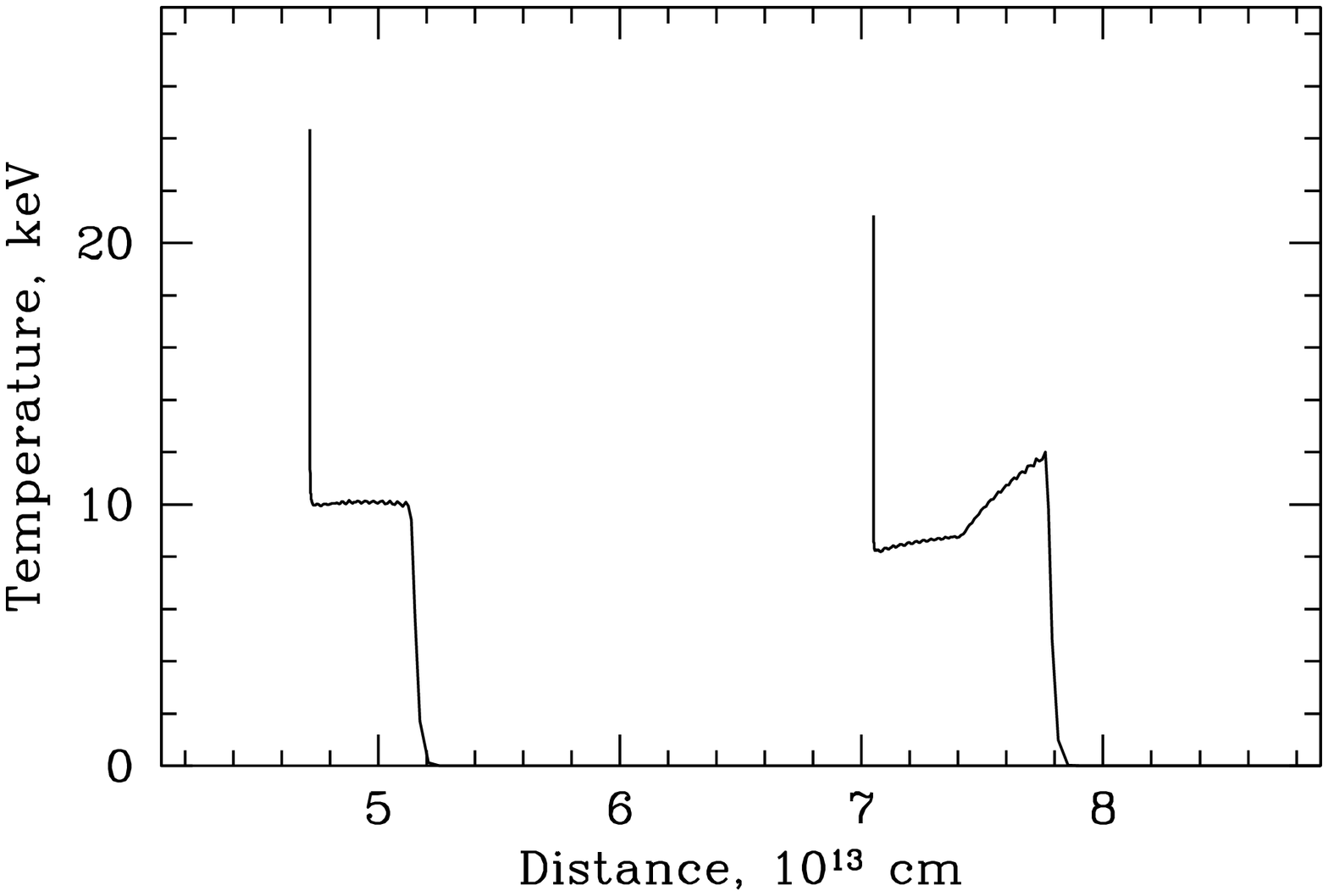}
\includegraphics[width=0.9\columnwidth,angle=0.,bb=20 145 570 500]{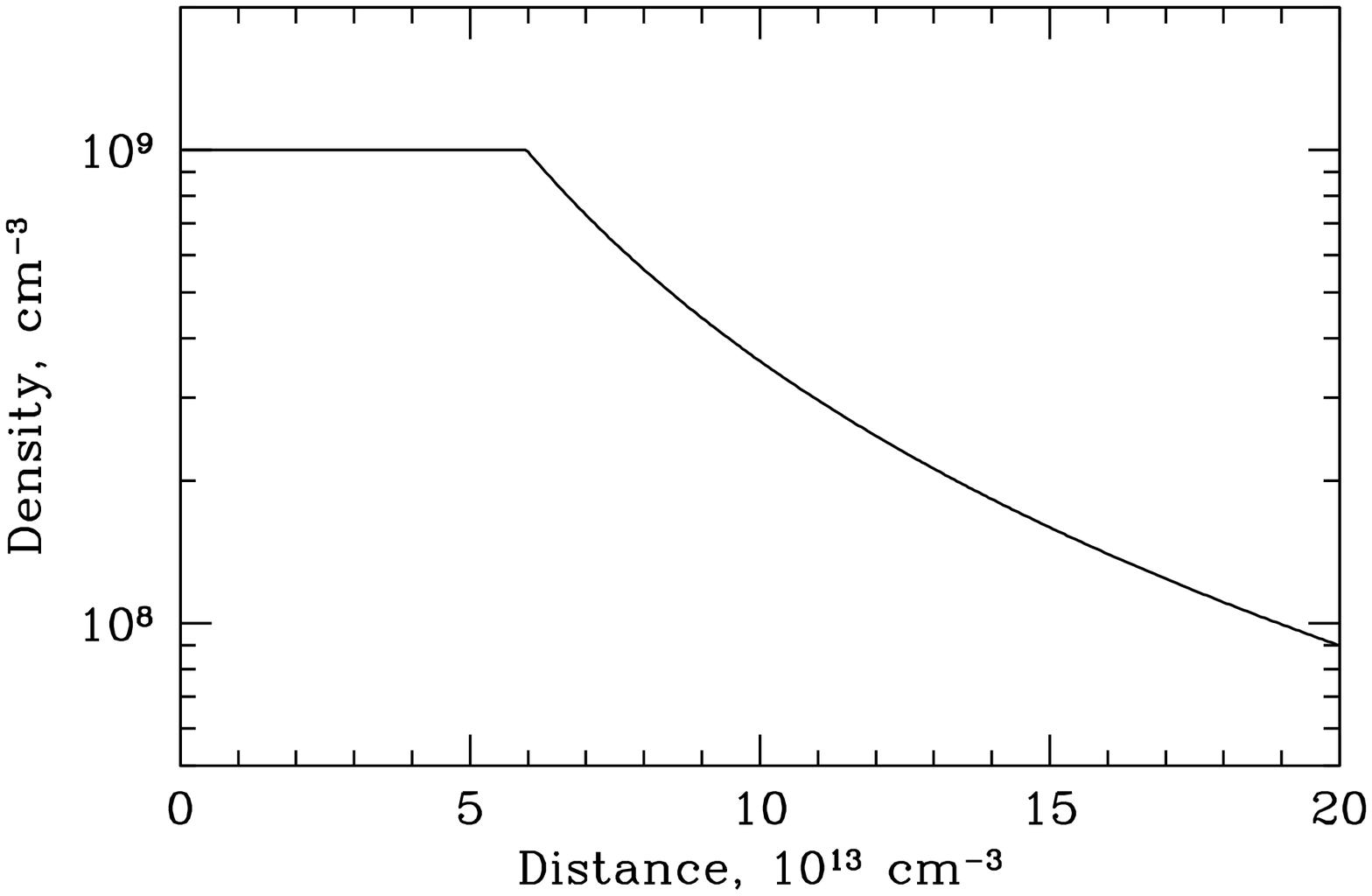}
\caption{Left: stellar wind temperature profiles when 
the shock moves with a constant velocity in a region 
with a constant density
of the interstellar medium (two days after the explosion) 
and in a region where the density decreases as $n \propto r^{-2}$ 
(three days after the explosion). 
Right: initial density profile of 
the interstellarmedium.} \label{tempprof_roperem_uconst}
\end{figure}

\begin{figure}[]
\includegraphics[width=0.9\columnwidth,angle=0.,bb=20 145 570 500]{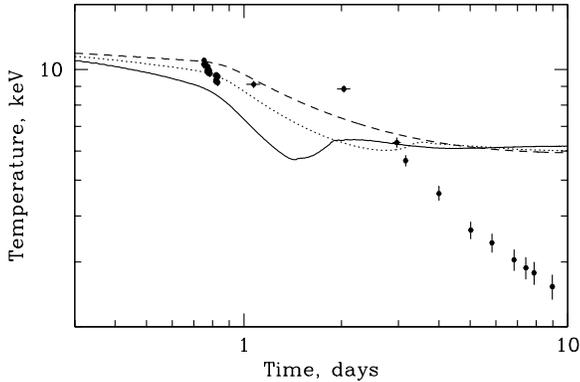}
\caption{Time dependences of the mean temperature at various 
stellar wind densities near the white dwarf: $10^9 cm^{-3}$ 
(dashed line), $5\times10^9 cm^{-3}$ (dotted line), 
and $10^{10} cm^{-3}$ (solid line). 
The piston velocity was kept constant ($U = 2700$ km/s).
The boundary of the transition from a constant 
density to a decreasing one is $r_c = 1.9\times10^{13}$ cm. 
The dots indicated the
observational data.} \label{temp_uconst_difn}
\end{figure}

\section{Rise Phase Of The Light Curve}

In Fig. 1(left), the dashed line indicates the fit to the
increase in observed (3-20 keV) flux by the model
$\propto t^3$. The observed rise in the source's 
luminosity is
well described by the law 

$L_{3-20keV} \sim (1.1\pm0.1)\times10^{22} t_{s}^3 d_{2kpc}^2$, 
where $d_{2 kpc}$ = d/2 kpc.

It follows from our calculations that as long as the
envelope moves with a constant velocity in a medium
with a constant density and the radiative cooling
of the heated matter is negligible, the post-shock
temperature remains approximately constant.Consequently,
the flux increases just as the hot-plasma
emission measure, 
$L \propto n^2V \propto r^3$, which for a constant
piston (and, hence, shock) velocity gives
$$
L=\Lambda n^2 {4\pi\over{3}} (D^3-U^3)t^3,
$$
where $D$ is the shock velocity and $U$ is the piston
velocity.

Thus, it can be said that the observed behavior of
the source's luminosity suggests that the shock in the
binary system moved through matter with a constant
density in this period. Consequently, we can estimate
the mean density of the interstellar medium near the
white dwarf. At temperature $kT = 10$ keV, the plasma
emissivity is $\Lambda_{3-20 keV} = 3.4\times10^{-24} erg\, cm^3 s^{-1}$
(the value was reduced to the total particle number
based on the adopted abundance) and the theoretical
time dependence of the luminosity is 
$L_{3-20\,keV}=93\,n_{1}^2\, U_{2700}^3\, t_{s}^3$, 
where $n_1$ is in $cm^{-3}$ and 
$U_{2700} =U/2700$ km/s. It thus follows that
$$
n_0(r<r_c) \sim 8.6\times 10^{9} d_{2kpc} U_{2700}^{-3/2} cm^{-3}.
$$

For the distance range 1.1-1.9 kpc measured by Barsukova
et al.(2006), the possible range of densities
near the white dwarf is $(4.8-8)\times10^9 cm^{-3}$.

The time elapsed from the outburst onset to the instant
at which the source reaches its peak luminosity
allows the distance to which the stellar wind density
remains approximately constant to be estimated: 
$r_c = D t_{peak} \sim 1.9\times10^{13}(t_{peak}/0.75 day)D_{3000}$, 
where $D_{3000} = D/3000$ km/s. This value agrees with $r_c$
obtained by analyzing the binary's sizes 
(see the section
above) at an orbital inclination $\sin i \sim0.3-0.4$.

\section{Mean Temperature Of The Emitting
Matter And Shock Velocity At Late
Expansion Phases}

\begin{figure}[]
\includegraphics[width=0.9\columnwidth,angle=0.,bb=20 145 570 500]{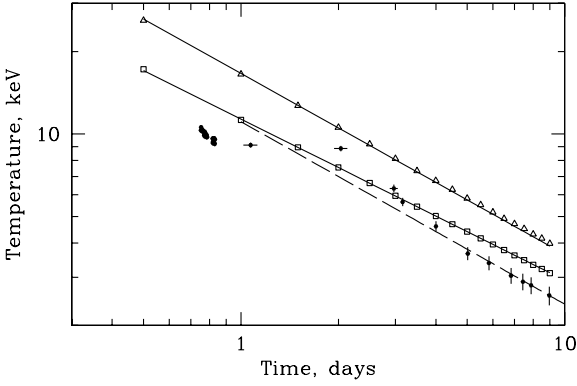}
\caption{Time dependences of the post-shock plasma 
temperature (open triangles; the solid line indicates 
the law $T\sim t^{-0.66}$)
and the mean radiation temperature (open squares; 
the solid line indicates the law $T \sim t^{-0.6}$) 
when the shock moves in the
Sedov phase in a medium with a density 
$n \propto r^{-2}$. The circles indicate 
the observational data. The dashed line indicates the
dependence $T_{keV} \sim 11.1t^{-0.66}_{days}$.}\label{temp_sedov_ror2}
\end{figure} 

The radiation temperature begins to decrease
$\sim$2 days after the outburst onset (Fig.1, right). 
The temperature
can decrease for several reasons:
\begin{itemize}

\item as a result of the shock passage into a region
with a decreasing density of the interstellar medium
at a constant velocity of the piston envelope;
\item as a result of the deceleration of the piston
envelope. In this case, the shock passes to the Sedov
phase (the regime of shock motion after an instantaneous
point explosion).
\end{itemize}

\subsection{Shock Motion Through a Region with a Decreasing
Density}

When the envelope moves with a constant velocity
and as the shock passes from a medium with a
constant density to a medium with a decreasing density,
the behavior of the emitting plasma temperature
changes. Although the post-shock plasma temperature
increases, since the shock velocity in a medium
with a density profile $n \propto r^{-2}$ 
depends on the piston
velocity as $D = 1.19U$ (Sedov 1945; Parker 1963),
the mean radiation temperature will decrease due to
adiabatic cooling. Typical post-shock plasma temperature
profiles for negligible radiative losses in this
case are shown in Fig. 15(left). The initial stellar wind
density profile is shown in Fig.15(right).

We calculated the motion of the piston envelope
with a constant velocity through a medium with the
density profile shown in Fig. 15 but fixed the radius of
the transition from a constant density to a decreasing
one at $r_c = 1.9\times10^{13}$ cm. Since the distance to the
source is uncertain, the stellar wind density near the
white dwarf estimated in Section ``Rise Phase Of The Light Curve'' 
is not the ultimate
value. Therefore, below we considered several
values of $n_0$: $10^9, 5\times10^9,$ and $10^{10} cm^{-3}$. 
The time
dependences of the temperature of the emitting matter
derived in these calculations are shown in Fig. 16.

We see from the figure that as a result of the piston
motion with a constant velocity in such a stellar
wind density profile, the mean temperature after some
decrease again reaches a constant value at late expansion
phases. Thus, the model with a constant envelope
velocity does not allow the observed behavior
of the temperature to be described at late expansion
phases. Consequently, the piston envelope should begin
to decelerate by this time, while the shock will
pass to the Sedov phase.

It also follows from Fig.16 that strong radiative
cooling (the curves for $n_0 = 10^{10}cm^{-3}$) is in conflict
with observations even for such a density profile.

\subsection{Shock Motion in the Sedov Regime}

In the Sedov regime, the shock motion ceases
to depend on the motion of the piston envelope: the
shock velocity decreases with time as $D \sim t^{-3/5}$ for
a uniform density distribution of the ambient medium
and as $D \sim t^{-1/3}$  for a decreasing ($n \propto r^{-2}$) density
of the medium(Sedov 1945). Consequently, the postshock
plasma temperature depends on time as 
$T \sim t^{-6/5}$ and $T \sim t^{-2/3}$, 
respectively, provided that the
radiative losses of the post-shock matter are negligible.

The calculated time dependences of the mean
temperature of the heated matter and the temperature
at the shock for the shock motion in the Sedov
phase in a medium with a decreasing density, 
$n_0=5\times10^{10} (r/2.1\times10^{11})^{-2} cm^{-3}$, 
are shown in Fig. 17.
The explosion energy was set equal to $5.77\times10^{40}$ erg.
The exponent in the time dependence of the mean
temperature $T \sim t^{-0.6}$ is close to both the exponent
for the temperature at the shock,  $T \sim t^{-0.66}$, and to
the observed law $T \sim t^{-0.7-0.6}$ on days 4-10 after
the explosion. Consequently, we may conclude that
the shock in CI Cam passed to the Sedov phase on
days 4-10 after the onset of envelope expansion.

The explosion energy can be estimated by comparing
the observed and model time dependences of
the mean temperature of the emitting matter in this
regime. For the time being, by the explosion energy
we mean the kinetic energy of the ejected envelope,
while below we will refine this concept. The shock velocity
in the Sedov phase in a medium with a density
$n_0 = A_1/r^2$ is
$$
D={{2}\over{3}}\left({{E}\over{A\alpha}}\right)^{1/3} t^{-1/3}.
$$ 
Here, $E$ is the explosion energy, $\alpha=2.1$ (Book 1994), 
and the coefficient
$A = A_1\mu m_p \sim 10^{11}n^9 r^2_{c,13} g\, cm^{-1}$, 
where $n_9 =n_0/10^9 cm^{-3}$ and $r_{c,13} = r_c/10^{13 }$ cm.

The time dependence of the temperature at the
shock is
$$
T_{keV}=1.6 \times 10^{-20}\left({{E_{erg}}\over{A}}\right)^{2/3} t_{days}^{-2/3}.
$$

It follows from the calculations presented in this
section that the flux-averaged temperature of the
post-shock matter is lower than that at the shock
itself, but their ratio has the same value of 1:1.3
(here, we assume that the radiative losses in the
plasma are negligible). The observed time dependence
of the radiation temperature on the fourth day
after the explosion can be fitted by the law 
$T_{keV} \sim 11.1t^{-2/3}_{days}$ 
(indicated in Fig. 17 by the dashed line).
Consequently, the required explosion energy is
$$
E\sim 2.65 \times10^{42} n_{9}\, r_{c,13}^2 erg.
$$

\section{Estimating The Mass Of The Escaped
Envelope. Description
Of The Observed Temperature
And Luminosity}

\begin{figure}[]
\includegraphics[width=0.9\columnwidth,angle=0,bb=20 145 570 500]{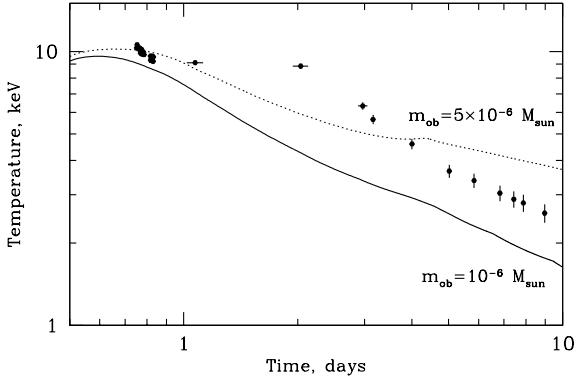}
\caption{Time dependences of the mean temperature for the 
expansion of envelopes with various masses. The envelope
expanded freely from the very beginning of the outburst (see the text).}\label{temp_difmas}
\end{figure}

In the previous sections, we showed the following:
\begin{itemize}
\item a constant mean temperature of the emitting
matter within the first 1-2 days after the onset of
envelope expansion can be obtained when the piston
envelope moves with an approximately constant velocity;
\item at late explosion phases (on days $\sim$4-10 after
the explosion onset), the shock motion in the stellar
wind passes to the Sedov regime.
\end{itemize}

The shock passage to the Sedov regime takes
place when the piston envelope transfers a substantial
fraction of its kinetic energy to the surrounding matter
and, consequently, decelerates greatly. Assuming
a uniform distribution of interstellar matter with a
density $n_0$, we can estimate, to a first approximation,
how the time $\tau_S$ of shock passage to the Sedov phase
depends on the envelope mass:
$$
M_{ej} \sim {4\over{3}}\pi (D \tau_{\rm S})^3 \mu m_p n_0,
$$
i.e.
$$
\tau_{S}\sim1.7 M_{ej,-6}^{1/3}n_{5e9}^{-1/3} (D_{3000})^{-1}s,
$$

where $M_{ej,-6}=M_{ej}/10^{-6} \msun$ is the envelope mass,
$n_{5e9}=n_0/(5\times10^{9}) cm^{-3}$ is the unperturbed stellar
wind density.

However, first, it follows from our estimates obtained
in the Section ``Rise Phase of the Light Curve''
that the shock traverses the region with a constant
density within the first 0.75 day, the stellar matter
density then drops and, hence, the envelope deceleration
efficiency decreases; second, it must be known
how the mean temperature behaves as the envelope
decelerates. Therefore, we performed numerical calculations
for various envelope masses in which its
velocity was specified at the initial time; subsequently,
it expanded freely, i.e., the envelope changed its velocity
as a result of its interaction with the medium.
The stellar wind density near the white dwarf was
set equal to $n_0 = 5\times10^9 cm^{-3}$ and the radius of the
transition from a constant density to a decreasing
one was taken to be $r_c = 1.9\times10^{13}$ cm. It follows
from these calculations (Fig. 18) that an envelope
with a mass $M_{ej} > 5\times10^{-6}\msun$ needed for it to expand 
with a constant velocity in the initial period of its
expansion has no time to decelerate sufficiently at late
phases, while an envelope with a smaller mass begins
to decelerate immediately, causing the radiation
temperature to drop immediately after the explosion.
As an example, Fig. 18 shows the time dependences
of the mean temperature for envelope masses 
$M_{ej} = 5\times10^{-6}\msun$ and $M_{ej} = 10^{-6}\msun$.

Consequently, we may conclude that the envelope
velocity in the initial period of its expansion is maintained
not by its mass but by an additional pressure
from the white dwarf. Most likely, one might expect
this to be the radiation pressure of the matter that
continues to burn on the white dwarf surface.

For the radiation pressure to be able to effectively
keep the envelope expansion velocity constant
within the first 1.5-2 days, the optical depth of the
envelope must be much larger than unity. If the envelope
temperature is assumed to be $\sim10^5-10^6$ K,
then the absorption cross section will be determined
by the absorption of incompletely ionized heavy elements,
which increase significantly the absorption
cross section compared to that for the scattering by
free electrons. We cannot accurately estimate the absorption
cross section in the envelope matter, because
its chemical composition and ionization fraction are
unknown, but we can estimate the photon absorption
cross section required for an envelope with a mass
$M_{ej} = 10^{-6}\msun$ to become optically thin 1.5 days after
the explosion.If the envelope expands spherically
symmetrically, filling the volume with a uniform density
$n$, then
$$
M_{\rm ej}\sim {{4\pi}\over{3}} R^3 n \mu m_p.
$$

Since $\tau \sim n \sigma R\sim 1$, we obtain:

$$
M_{\rm ej}\sim {{4\pi}\over{3 \sigma}} R^2 \mu m_p.
$$

Moving with the velocity $U = 2700$ km/s, the envelope
will traverse a distance $R\sim3.5\times10^{13}$ cm
within 1.5 days and will become optically thin if the
photoabsorption cross section in the matter is 
$\sigma \sim 2\times10^{-24} cm^2$.

\subsection{Best-Fit Models for the Observed Time Dependence
of the Radiation Temperature}

\begin{figure*}[]
\centerline{
\hbox{
\includegraphics[width=0.9\columnwidth,angle=0,bb=20 145 570 500]{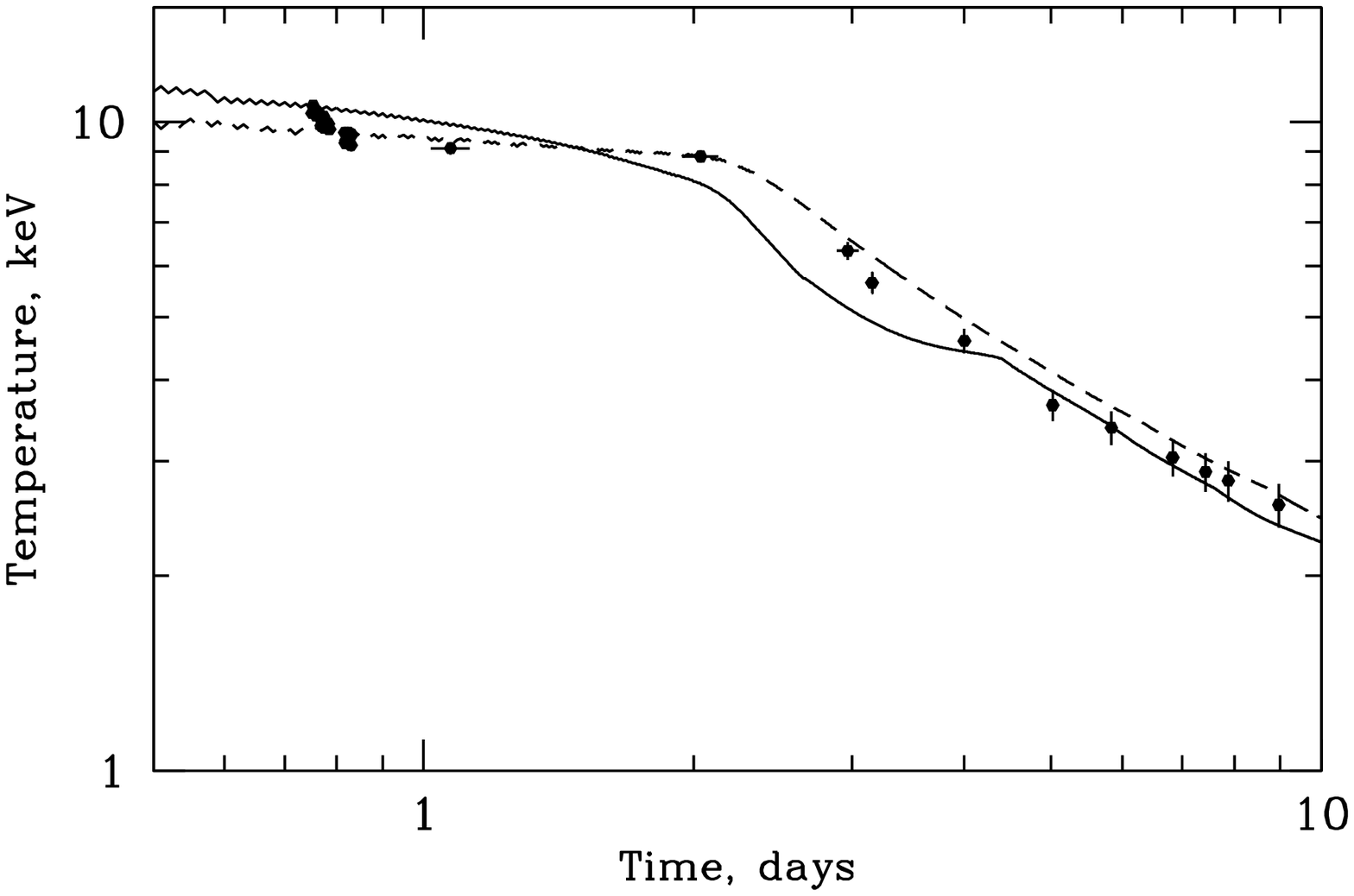}
\includegraphics[width=0.9\columnwidth,angle=0,bb=20 145 570 500]{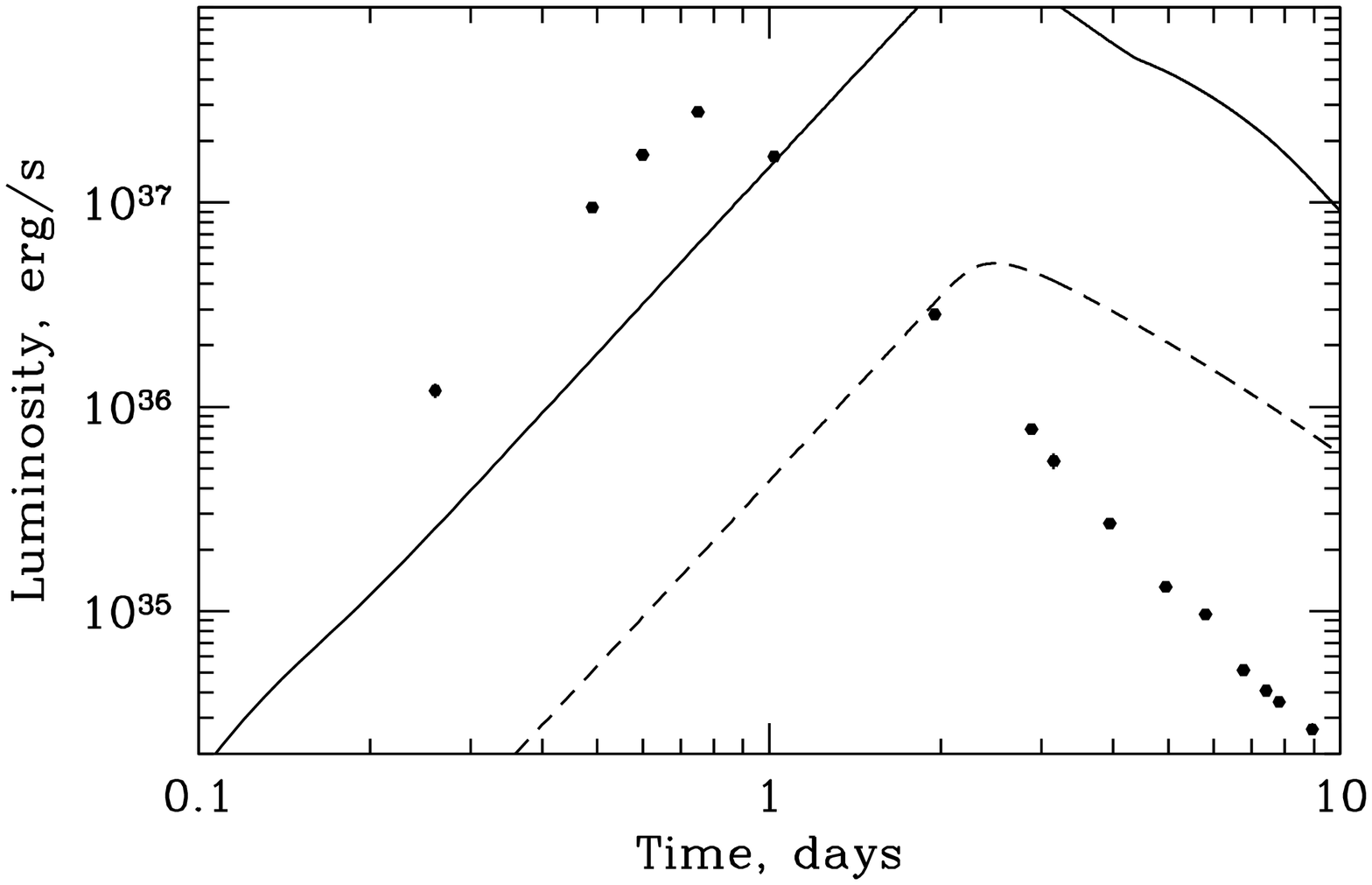}
}
}
\caption{Left: time dependences of the mean temperature 
when the envelope expands in a medium with $r_c = 5.2\times10^{13}$ cm. 
The solid and dashed lines indicate the dependence 
for $n_0 = 5\times10^9 cm^{-3}$ and $n_0 = 10^9 cm^{-3}$, 
respectively. Right: time dependence
of the luminosity.}\label{temp_lum_gr5e13}
\end{figure*}

To describe best the observed time dependence
of the mean temperature, we must assume that the
mean stellar wind density around the white dwarf
remains constant up to the distance that the shock
traverses in $\sim$2 days, i.e., 
$r_c = 5.2\times10^{13}D_{3000}$ cm.
Figure 19 shows the best-fit models for 
$r_c = 5.2\times10^{13}$ cm and two matter densities near the white
dwarf: $n_0 = 10^9 cm^{-3}$ (dashed line) 
and $n_0 = 5\times10^9 cm^{-3}$ (solid line). 
However, without describing
these models in detail, we will immediately say that
such a value of $r_c$ is possible at low orbital inclinations
($\sin i < 0.1$). In addition, the main shortcoming
of these models is that such a density distribution
causes the X-ray luminosity to increase as 
$L \propto t^3$
until approximately the same time (2 days since the
onset of envelope expansion).This is in conflict with
the observational data (Fig.19,right) suggesting that
the luminosity reaches its peak $\sim$0.75 day after the
outburst onset. Therefore, below we assume that the
region with an approximately constant stellar wind
density around the white dwarf extends to 
$r_c = 1.9\times10^{13}$ cm; 
subsequently, the density falls as $n \propto r^{-2}$.

As a result of the uncertainty in the stellar wind
density $n_0$ near the white dwarf and the uncertainty
in the time of envelope motion under the action of an
external force, several models with different envelope
masses describe the time dependence of the mean
temperature equally well (Fig. 20). The model parameters
are given in the table; as the initial conditions,
we specified the stellar wind density near the white
dwarf $n_0$, the envelope mass $M_{ej}$, and the envelope
pushing time $\Delta t$ in our calculations. 
Also given in the
table are the kinetic energy of the envelope $E_{ej}$ and
the total energy of the post-shock stellar wind $E_{sw}$
at the time when the external force ceases to act
and a theoretical estimate of the explosion energy via
the observed temperature at the envelope deceleration
phase $E_{sedov}$. In these models, by the explosion energy
we mean the kinetic energy of the envelope and the total
energy of the post-shock matter at the time when
the envelope ceases to be affected by the external
force.The envelope velocity in the initial period of its
expansion was $U = 2700$ km/s.

We see from the table that the kinetic energy of the
envelope in model 1 is lower than the total energy of
the post-shock matter; accordingly, it is not the main
source of the shock energy at late phases. Since a
decrease in the envelope mass in this model will cause
the mean temperature in the Sedov regime to change
only slightly, the derived envelope mass is most likely
an upper limit in this case.

It should be noted that although the shock passed
to the deceleration phase at late expansion stages in
models 1 and 3, this is not yet the Sedov regime,
when the post-shock temperature depends on time as
$\sim t^{-2/3}$.

We see from Fig. 20 that there are no qualitative
differences in the description of the time dependences
of the mean temperature by these models.Nevertheless,
a small envelope mass makes it difficult for the
envelope to move under the action of an external force
within 1.5 days, because it rapidly becomes optically
thin. Therefore, we will consider the range of possible
envelope masses $M_{ej} \sim 10^{-7}-10^{-6}\msun$.

\begin{figure}[]
\includegraphics[width=0.9\columnwidth,angle=0,bb=20 145 570 500]{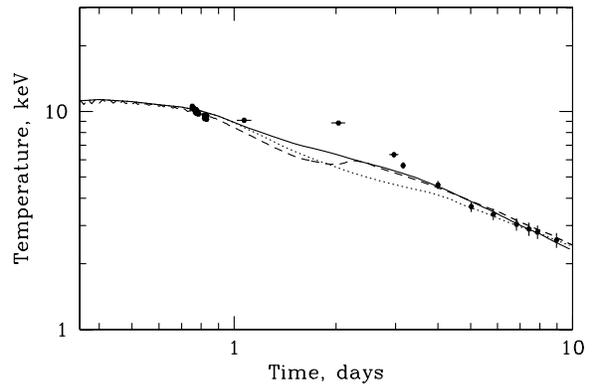}
\caption{Time dependences of the mean temperature for 
models with various initial conditions: model 1 (solid line), model 2
(dotted line), and model 3 (dashed line). The models 
are described in the text. The dots indicate the observational data.}\label{temp_final}
\end{figure}

\subsection{Accelerated Motion of the Envelope}
\begin{figure}[]
\includegraphics[width=0.9\columnwidth,angle=0,bb=20 145 570 500]{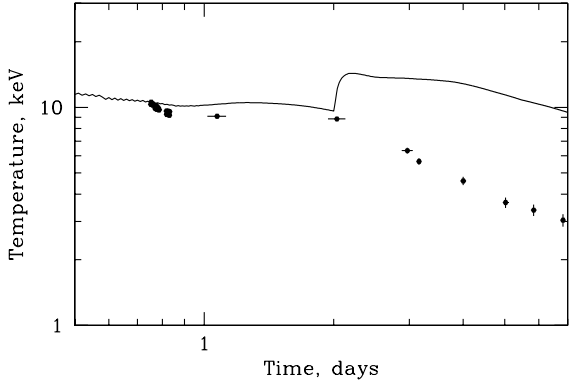}
\caption{Time dependences of the mean temperature for
accelerated motion of the envelope (see the text).T he dots
indicate the observational data.}\label{temp_ac}
\end{figure}

The shock passage into the region with a decreasing
density $\sim$0.75 day after the outburst causes a
decrease in the mean temperature and a discrepancy
between the calculated and observed temperatures on
the second day of envelope expansion. In order to
try to remove this discrepancy, we performed calculations
with accelerated envelope motion. It follows
from these calculations that for the temperature to be
kept constant within the first two days, the envelope
velocity must be increased by a factor of $\sim$2 by the
second day of its expansion, but as a result of this
measure, the calculated mean temperature turns out
to be much higher than the observed one at late
expansion stages. In Fig. 21, the solid line indicates
the time dependence of the mean temperature for the
model with the following parameters: the stellar wind
density near the white dwarf was $n_0 = 5\times10^9 cm^{-3}
(r_c = 1.9\times10^{13} cm$); the envelope velocity was kept
constant and equal to $U = 2700$ km/s within the
first 0.8 day; subsequently, the velocity increased in
such a way that it was twice that at the onset of
acceleration by the second day; after the second day,
the envelope expanded freely, with the envelope mass
after the onset of its free expansion having no effect
on the shock motion. It follows from these calculations
that the absence of envelope acceleration in our
calculations is not responsible for the discrepancy in
temperatures on the second day of expansion.

\subsection{Description of the Binary Luminosity
During its Outburst}

\begin{figure}[]
\centerline{
\includegraphics[width=0.9\columnwidth,angle=0,bb=20 145 570 500]{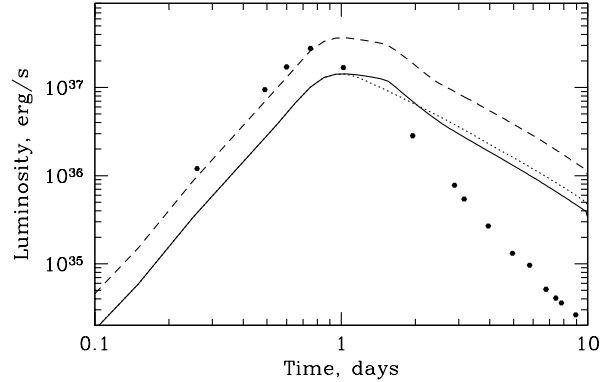}
}
\caption{Time dependences of the luminosity for various
initial conditions: model 1 (solid line), model 2 (dotted
line), and model 3 (dashed line).The dots indicate the
observational data.}\label{lum_final}
\end{figure}

\begin{figure}[]
\centerline{
\includegraphics[width=0.9\columnwidth,angle=0,bb=20 145 570 500]{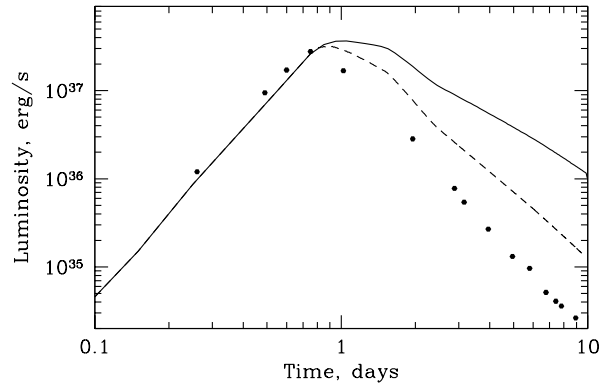}
}
\caption{Time dependences of the system's luminosity 
for a spherically symmetric stellar wind (solid line) 
and a stellar wind
with a dense disk of constant thickness (dashed line). 
The dots indicate the observed luminosity}\label{lum_disk}
\end{figure}

The time dependences of the luminosity for models
1, 2, and 3 are shown in Fig. 22. The discrepancy
between the calculated and observed luminosity at the
rise phase is unimportant, since there is an uncertainty
in the estimated stellar wind density near the
white dwarf due to the uncertainty in the distance
to the system. At the same time, the discrepancy in
the time dependences of the luminosity at the decline
phase can be explained by the fact that the behavior
of the luminosity during the outburst is subjected
to density variations much more strongly than the
behavior of the radiation temperature. Therefore, incomplete
agreement between the model and observed
time dependences of the luminosity during the 1998
outburst of CI Cam is most likely related to peculiarities
of the actual stellar wind density distribution
in the binary system, which we do not know
at present and which cannot be taken into account
in spherically symmetric calculations. For example,
one might expect the stellar wind density distribution
around the B[e] star to be nonspherically symmetric
and more closely resemble a disk. In this case, the
heated matter of the stellar wind disk will make the
main contribution to the observed X-ray emission. A
disk with a constant opening angle will only reduce
the luminosity normalization but will not change the
pattern of luminosity behavior qualitatively. However,
if the disk has a constant thickness (see, e.g., Ignace
et al. 2004), then the decline in luminosity during the
outburst will be faster.

Figure 23 shows the time dependences of the
system's luminosity for the shock motion in a medium
with a spherically symmetric density distribution
around the white dwarf and under the assumption
that dense matter in the form of a disk with a
constant thickness makes the main contribution to
the emission at the decline phase of the light curve.
The luminosity at the decline phase for a disk wind
in this figure was estimated only qualitatively (the
normalization of the light curve at the decline phase
was chosen in such a way that the luminosity at the
peak of the light curve coincided with its value from
the light curve at the rise phase) by assuming that the
time dependence of the temperature did not change
compared to the spherically symmetric case.

\begin{table*}\label{models}
\caption{Parameters of the best-fit models for the observational data}
\centering
\begin{tabular}{c|c|c|c|c|c|c}
&&&&&\\
Model &$\Delta t$, days&$n_0$, $10^9$ cm$^{-3}$&$M_{ej}$, $ \msun$& $E_{ej}$,$10^{43}$ erg&$E_{sw}$,$10^{43}$ erg&$E_{sedov}$,$10^{43}$ erg\\
number&&&&&&\\
\hline
1&1.5&5&$2.85\times10^{-7}$&2&3.3&4.8\\
2&1&5&$10^{-6}$&8&1.7&4.8\\
3&1.5&8&$10^{-6}$&8&5&7.7\\
\end{tabular}
\end{table*} 

\section{Contribution Of The Matter Heated
By The Reverse Shock
To The Observed Temperature
And Luminosity}

\begin{figure}[]
\centerline{
\includegraphics[width=0.9\columnwidth,angle=0,bb=20 145 570 500]{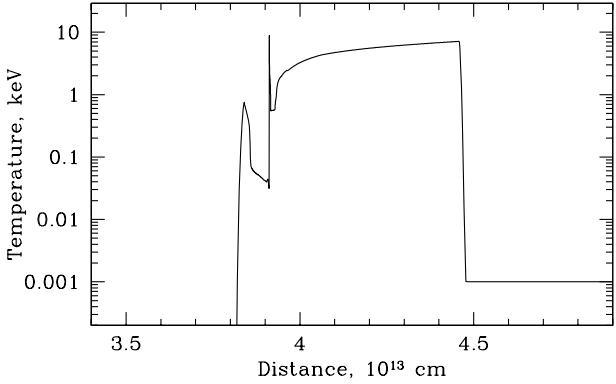}
}

\caption{Temperature profiles downstream of the forward
and reverse shocks.The forward shock is at
$r \sim 4.5\times10^{13}$ cm, the reverse shock 
is at $r\sim3.8\times10^{13}$ cm,
and the contact boundary is at 
$r\sim3.9\times10^{13}$ cm.}\label{tempprof_obolochka}

\end{figure}

The time dependence of the temperature downstream
of the reverse shock (RS) is affected by many
factors that are not known exactly to date, such as the
density profile in the envelope; whether the envelope
expands, having a finite thickness, or continuously
fills the entire space up to the white dwarf.
In this paper, we will give only qualitative estimates
based on the decay of an arbitrary discontinuity (Riemann problem)
in the plane case (Rozhdestvensky and Yanenko
1978). These will show that the contribution from
the post-RS matter to the observed emission may
be neglected, to a first approximation. Assuming the
expansion of the envelope to be uniform, its density
depends on the radius as
$$
n_{ej}={{M_{ej}}\over{4/3\pi r^3 \mu m_p}}=3.7\times10^{15}M_{-6,ej} r_{5\times10^{11}}^{-3}.
$$

Under certain conditions, a reverse rarefaction
wave can be generated in the envelope at the initial
time, which will subsequently turn into a shock as
the envelope expands. Let us estimate the distance
at which the RS generation condition is satisfied for
the derived envelope velocity of 2700 km/s. The
dependence of the velocity of the envelope at which
the RS will travel into it on the ratio of the pressures
in the envelope $p_{ej}$ and the ambient medium $p_0$ is
$$
U_{shock}=(1-h)c_0\left(\sqrt{{p_{ej}/p_0+h}\over{1+h}}-\sqrt{{1+h}\over{p_{ej}/p_0+h}}\right),
$$
where  $h={{\gamma-1}\over{\gamma+1}}$. For the adopted stellar
wind temperature $kT_0 = 1$ eV, the speed of sound
is $c_0 = 16$ km/s. For our estimates, we will take
the stellar wind density $n_0 = 10^{10} cm^{-3}$. 
For adiabatic
envelope expansion, the pressure ratio depends
on the radius as 
$p_{ej}/p_0=n_{ej,0}/n_0(5\times10^{11}/r)^{3\gamma}=3.7\times10^5r_{5\times10^{11}}^{-5},$
where $n_{ej,0}$ corresponds to the envelope
density at $r = 5\times10^{11}$ cm; the envelope and
stellar wind temperatures at this radius were also
assumed to be equal.

Assuming that $p_{ej}/p_0 >> 1$, we find
$$
p_{ej}/p_0=\left({{U_{shock}}\over{c_0(1-h)}}\right)^2(1+h)\sim6.3\times10^4.
$$

The distance that the envelope shoul reach in order
to have such a density (for the $M_{ej}=10^{-6} \msun$) is $r = 7\times10^{11}$ cm , 
i.e., the RS
is generated $\sim$0.7 h after the onset of expansion.
As the envelope expands, the pressure decreases,
causing the RS to strengthen. By the shock strength
we will mean the ratio of the shock velocity to the
speed of sound in the medium. Let us estimate the
time when the RS will be able to heat the matter up to
1 keV. The RS strength $M_r$ depends on the strength
of the forward shock (FS) $M_f$ as
$$
M_{r}^2=\alpha M_{f}^2+\beta,
$$
where $\alpha = p_0/p_{ej}$ and  
$\beta={{h}\over{1+h}}{{p_{ej}-p_0}\over{p_{ej}}}$
if the adiabatic
indices of the matter in the ambient medium
and the envelope are equal; 
 $M_{r}={{D_{ej}}\over{c_{ej}}}$,  $M_{f}={{D}\over{c_0}}\sim187.5$,  
$D_{ej}=|D_1-U_1|$ is the RS velocity in the
envelope matter frame of reference.

For the RS to be able to heat the matter to 1 keV,
its velocity relative to the envelope must be 
$D_{ej} \sim 915$ km/s. It follows from the relationship between
the shock strengths that
$$
{{D^{2}_{ej}}\over{c_{ej}^2}}\sim{{p_0}\over{p_{ej}}}M^2_f
$$

The pressure in the envelope can be expressed
in terms of $p_{ej}=\gamma^{-1} \mu m_p n_{ej}c_{ej}^2$. 
Thus , for the RS to
be able to heat the matter to 1 keV, the density in
the envelope must be $n_{ej} = 1.1\times10^{11} cm^{-3}$.
Consequently, the envelope must expand to the distance
$r = 1.6\times10^{13}$ cm, i.e., $\sim$0.7 day should be elapsed
since the onset of expansion.
Let us estimate the transparency of the post-
RS matter on 0.7 day after the onset of expansion
if the main mechanism responsible for the matter
opacity is Thomson scattering: 
$\tau\sim 4\times n_{ej} \sigma_T\Delta r \sim 4\times 1.1\times10^{11}\times 6.65\times10^{-25}\times 3\times10^{12} \sim 0.9$. 
For the absorption cross section $\sigma=2\times10^{-24} cm^2$ obtained
above, $\tau\sim2.6$. As long as the heated post-RS
region is optically thick, the entire envelope emits
the Eddington luminosity and the temperature of the
emergent radiation is very low, $\sim3$ eV.

The envelope becomes optically thin as it expands.
To demonstrate the behavior of the post-RS temperature,
Fig. 24 shows an example of the temperature
profile in the case where the envelope becomes optically
thin almost immediately and cools down through
the radiation of an optically thin plasma (the FS is at
$r \sim 4.5\times10^{13}$ cm, the RS is at $r\sim3.8\times10^{13}$ cm,
and the contact boundary is at $r\sim3.9\times10^{13}$ cm). 
It follows from this profile that as long as the envelope
density was high (the region formed near the contact
discontinuity), the temperature of the post-RS matter
was low, since it effectively cooled down through
radiation, but the post-RS matter temperature began
to increase as the envelope expanded and became
rarefied. We calculated the mean temperature of the
radiation coming only from the FS-heated matter and
the mean temperature of the entire emitting matter
for this temperature profile. As we suggested, the
contribution from the post-RS matter does not affect
the mean temperature and, hence, it will not affect the
luminosity in the observed 3-20 keV energy band.

Thus, our preliminary estimates show that, to a
first approximation, the contribution of the radiation
from the RS-heated matter to the observed 3-20 keV
luminosity and the mean temperature is small compared
to that of the radiation from the FS-heated
matter.

\section{Conclusions}

Using analytical estimations and numerical simulations,
we were able to describe the intense X-ray
outburst in the binary system CI Cam observed
in 1998 in terms of the model of the interaction between
the expanding envelope ejected as the result of
a thermonuclear explosion on the white dwarf surface
(classical nova explosion) and the stellar wind from
the optical component. In particular, we showed the
following:

\begin{itemize}
\item The mean radiation temperature of the stellar
wind matter passed through the shock differs significantly
from that of the matter immediately behind
the shock front. Therefore, numerical calculations are
needed to properly determine the classical nova envelope
expansion parameters.

\item According to our preliminary estimates, the
contribution from the reverse shock wave heated matter to the observed
mean temperature and the 3-20 keV luminosity
is smaller than that from the forword shock wave heated matter.

\item The radiation temperature measured at early
envelope expansion phases of the classical nova allowed
its expansion velocity to be estimated:
$\sim$2700 km/s.

\item In our model, the envelope is ejected from
the white dwarf as the result of explosive thermonuclear
burning. It has an expansion velocity of
$\sim$2700 km/s already on 0.1-0.5 day after the
explosion onset and, being optically thick, moves under
the radiation pressure from the white dwarf with a
constant velocity for the first $\sim$1-1.5 days. Thus,
we have been able to measure the envelope velocity
almost immediately after the explosion for the first
time.

\item Subsequently, the envelope probably becomes
optically thin and decelerates while interacting with
the stellar wind from the optical component. The
shock in the stellar wind passes to the Sedov phase
with the time dependence of the mean temperature
$T\sim t^{-0.7-0.6}$.

\item By comparing the observed rise in luminosity
with the theoretical dependence, we estimated the
stellar wind density near the white dwarf, 
$n_0(r < r_c) \sim 8.6\times10^9d_{2 kpc}U^{-3/2}_{2700} cm^{-3}$. 
In the simplest
model of the stellar wind density distribution, this corresponds
to a stellar mass loss rate in the stellar wind
of $\sim(1-2)\times10^{-6}\msun/yr$. This estimate agrees well
with the values obtained by Robinson et al. (2002).

\item The observed time dependence of the temperature
of the emitting matter at late envelope expansion
stages allowed the mass of the ejected envelope to be
constrained in our model, $\sim10^{-7}-10^{-6}\msun$.

\end{itemize}

\begin{acknowledgements}
This work was supported by the Russian Foundation
for Basic Research (project no. 07-02-01051),
NSh-5579.2008.2, the Foundation for Support of
Russian Science, the ``Origin and Evolution of Stars
and Galaxies'' Program of the Presidium of the
Russian Academy of Sciences, and Priority Program
no.1177 (``Witnesses of the History of Space: the
Formation and Evolution of Galaxies, Black holes,
and Their Environment''). E. V. Filippova also thanks
D.Docenko and D.Giannios for helpful discussions
and valuable remarks.
\end{acknowledgements}

\section*{References}

1. E. A. Barsukova, N. V. Borisov , A. N. Burenkov, et al.,
Astron.Rep. 50, 664 (2006).

2. G. T. Bath and G. Shaviv, Mon.Not.R.Astron.Soc.
183, 515 (1978).

3. M. F. Bode and F. D. Kahn, Mon.Not.R.Astron.Soc.
217, 205 (1985).

4. M. F. Bode, T. J. O'Brien, J. P. Osborne, et al., Astrophys.
J. 652, 629 (2006).

5. L. Boirin, A. N. Parmar, T. Oosterbroek, et al., Astron.
Astrophys. 394, 205 (2002).

6. D.~L. Book, Shock Waves 4, 1 (1994).

7. J. S. Clark, A. S. Miroshnichenko, V. M. Larionov,
et al., Astron. Astrophys. 356, 50 (2000).

8. R. Das, D. P. K. Banerjee, and N. M. Ashok, Astrophys.
J. 653, L141 (2006).

9. T. Dumm, D. Folini, H. Nussbaumer, et al., Astron.
Astrophys. 354, 1014 (2000).

10. G. B. Field, Astrophys.J. 142, 531 (1965).

11. J. S. Gallagher and S. Starrfield, 
Ann.Rev.Astron.Astrophys. 16, 171 (1978).

12. S. A. Glasner, E. Livne, and J. W. Truran, Astrophys.
J. 625, 347 (2005).

13. S. A. Glasner, E. Livne, and J. W. Truran, Astrophys.
J. 665, 1321 (2007).

14. N. Grevesse and A. Sauval, Space Sci.Rev. 85, 161
(1998).

15. R. I. Hynes, J. S. Clark, E. A. Barsukova, et al.,
Astron.Ast rophys. 392, 991 (2002).

16. R. Ignace, J. Cassinelli, and J. Bjorkman, Astrophys.
J. 459, 671 (1996).

17. M. Ishida, K. Morio, and Y. Ueda, Astrophys. J. 601,
1088 (2004).

18. H.-T. Janka, T. Zwerger and R. Moenchmeyer, Astron.
Ast rophys. 268, 360 (1993).

19. P. Kahabka and E. P. J. van den Heuvel, 
Ann.Rev.Astron.Astrophys. 35, 69 (1997).

20. M. Kato and I. Hachisu, Astrophys. J. 437, 802
(1994).

21. A. Kercek, W. Hillebrandt, and J. W. Truran, Astron.
Astrophys. 337, 379 (1998).

22. A. Kercek, W. Hillebrandt, and J. W. Truran, Astron.
Astrophys. 345, 831 (1999).

23. H. M. Lloyd, T. J. O'Brien, M. F. Bode, et al., Nature
356, 222 (1992).

24. J. MacDonald, M. Y.Fujimoto, and J. W. Truran,
Astrophys.J. 294, 263 (1985).

25.K. Masai, Astrophys. J. 437, 770 (1994).

26. A. J. Mioduszewski and M. P. Rupen, Astrophys.J.
615, 432 (2004).

27. K. Mukai and M. Ishida, Astrophys.J . 551, 1024
(2001).

28. W. F. Noh, Comp. Phys. 72, 78 (1978).

29. T. J. O'Brien, M. F. Bode, and F. D. Kahn, 
Mon.Not.R.Astron.Soc. 255, 683 (1992).

30. T. J. O'Brien, H. M. Lloyd, and M. F. Bode, 
Mon.Not.R.Astron.Soc. 271, 155 (1994).

31. A. Orr, A. N. Parmar, M. Orlandini, et al., Astron.
Astrophys. 340, L19 (1998).

32. E. N. Parker, Interplanetary Dynamical Processes
(Interscience, New York, 1963).

33. D.Prialnik, Astrophys.J. 310, 222 (1986).

34. D.Prialnik and A. Kovetz, AIP Conf.Proc. 797, 319
(2005).

35. M. G. Revnivtsev, A. N. Emel'yanov, and
K. N. Borozdin, Astron.Lett. 25, 294 (1999).

36. E. L. Robinson, I. I. Ivans, and W. F. Welsh, Astrophys.
J. 565, 1169 (2002).

37. B. L. Rozhdestvensky and N. N. Yanenko, Systems of
Quasilinear Equations and Their Applications to
Gas Dynamics (Nauka, Moscow, 1978; Amer.Math.
Soc., Providence, 1983).

38. A. A. Samarsky and Yu. P. Popov, Difference Solution
Methods of Gas-Dynamic Problems (Nauka,
Moscow, 1992) [in Russian].

39. L.I. Sedov, Prikl. Matem. Mekh. 9, 295 (1945).

40.L. I. Sedov, Similarity and Dimensional Methods
in Mechanics (Nauka, Moscow, 1981; CRC Press,
Boca Raton, 1993).

41. D. Smith, R. Remillard, J. Swank, et al., IAU Circ.
6855 (1998).

42. J. L. Sokoloski, G. J. M. Luna, K. Mukai, and
S. J. Kenyon, Nature 442, 276 (2006).

43. L. Spitzer, Physical Processes in the Interstellar
Medium (Wiley, New York, 1998; Mir, Moscow,
1981).

44. S. Starrfield, W. M. Sparks, and J. W. Truran, Astrophys.
J. 291, 136 (1985).

45. Y. Tanaka and N. Shibazaki, Ann.Rev.Astron.Astrophys.
34, 607 (1996).

46. Y. Ueda ,M. Ishida, H. Inoue, et al., Astrophys.J. 508,
L167 (1998).

47. R. Wagner and S. Starrfield, IAU Circ. 6857 (1998).

48. R. Walder, D. Folini, and S. Shore, Astron. Astrophys.
484, 9 (2008).

49. O. Yaron, D. Prialnik, M. M. Shara, and A. Kovetz,
Astrophys.J. 623, 398 (2005).

\end{document}